\documentclass[]{aastex631}

\accepted{\today}

\shorttitle{ALMA Observations of Comet 46P/Wirtanen}
\shortauthors{Cordiner et al.}

\begin{document}

\title{Gas Sources from the Coma and Nucleus of Comet 46P/Wirtanen Observed Using ALMA}

\correspondingauthor{M. A. Cordiner}
\email{martin.cordiner@nasa.gov}

%\author[0000-0001-8233-2436]{M. A. Cordiner et al.}

\author[0000-0001-8233-2436]{M. A. Cordiner}
\affiliation{Astrochemistry Laboratory, NASA Goddard Space Flight Center, 8800 Greenbelt Road, Greenbelt, MD 20771, USA.}
\affiliation{Department of Physics, Catholic University of America, Washington, DC 20064, USA.}
\author[0000-0002-6006-9574]{N. X. Roth}
\affiliation{Astrochemistry Laboratory, NASA Goddard Space Flight Center, 8800 Greenbelt Road, Greenbelt, MD 20771, USA.}
\affiliation{Department of Physics, Catholic University of America, Washington, DC 20064, USA.}
\author[0000-0001-7694-4129]{S. N. Milam}
\affiliation{Astrochemistry Laboratory, NASA Goddard Space Flight Center, 8800 Greenbelt Road, Greenbelt, MD 20771, USA.}
\author[0000-0002-2662-5776]{G. L. Villanueva}
\affiliation{Planetary Systems Laboratory, NASA Goddard Space Flight Center, 8800 Greenbelt Road, Greenbelt, MD 20771, USA.}
\author[0000-0002-8130-0974]{D. Bockel{\'e}e-Morvan}
\affiliation{LESIA, Observatoire de Paris, Universit{\'e} PSL, CNRS, Sorbonne Universit{\'e}, Universit{\'e} de Paris, 5 place Jules Janssen, F-92195 Meudon, France.}
\author[0000-0001-9479-9287]{A. J. Remijan}
\affiliation{National Radio Astronomy Observatory, Charlottesville, VA 22903, USA.}
\author[0000-0001-6752-5109]{S. B. Charnley}
\affiliation{Astrochemistry Laboratory, NASA Goddard Space Flight Center, 8800 Greenbelt Road, Greenbelt, MD 20771, USA.}
\author[0000-0003-2414-5370]{N. Biver}
\affiliation{LESIA, Observatoire de Paris, Universit{\'e} PSL, CNRS, Sorbonne Universit{\'e}, Universit{\'e} de Paris, 5 place Jules Janssen, F-92195 Meudon, France.}
\author[0000-0003-2414-5370]{D. C. Lis}
\affiliation{Jet Propulsion Laboratory, California Institute of Technology, 4800 Oak Drove Drive, Pasadena, CA 91109, USA.}
\author[0000-0001-8642-1786]{C. Qi}
\affiliation{Harvard-Smithsonian Center for Astrophysics, 60 Garden Street, MS 42, Cambridge, MA 02138, USA.}
\author[0000-0002-6391-4817]{B. P. Bonev}
\affiliation{Department of Physics, American University, Washington D.C., USA.}
\author[0000-0003-2414-5370]{J. Crovisier}
\affiliation{LESIA, Observatoire de Paris, Universit{\'e} PSL, CNRS, Sorbonne Universit{\'e}, Universit{\'e} de Paris, 5 place Jules Janssen, F-92195 Meudon, France.}
\author[0000-0002-1545-2136]{J. Boissier}
\affiliation{Institut de Radioastronomie Millimetrique, 300 rue de la Piscine, F-38406, Saint Martin d'Heres, France.}

\begin{abstract}

Gas-phase molecules in cometary atmospheres (comae) originate primarily from (1) outgassing by the nucleus, (2) sublimation of icy grains in the near-nucleus coma, and (3) coma (photo-)chemical processes. However, the majority of cometary gases observed at radio wavelengths have yet to be mapped, so their production/release mechanisms remain uncertain. Here we present observations of six molecular species towards comet 46P/Wirtanen, obtained using the Atacama Large Millimeter/submillimeter Array (ALMA) during the comet's unusually close ($\sim0.1$~au) approach to Earth in December 2018. Interferometric maps of HCN, CH$_3$OH, CH$_3$CN, H$_2$CO, CS and HNC were obtained at an unprecedented sky-projected spatial resolution of up to 25~km, enabling the nucleus and coma sources of these molecules to be accurately quantified. The HCN, CH$_3$OH and CH$_3$CN spatial distributions are consistent with production by direct outgassing from (or very near to) the nucleus, with a significant proportion of the observed CH$_3$OH originating from sublimation of icy grains in the near-nucleus coma (at a scale-length $L_p=36\pm7$~km). On the other hand, H$_2$CO, CS and HNC originate primarily from distributed coma sources (with $L_p$ values in the range 550--16,000~km), the identities of which remain to be established. The HCN, CH$_3$OH and HNC abundances in 46P are consistent with the average values previously observed in comets, whereas the H$_2$CO, CH$_3$CN and CS abundances are relatively low.

\end{abstract}

%% Keywords should appear after the \end{abstract} command. 
%% The AAS Journals now uses Unified Astronomy Thesaurus concepts:
%% https://astrothesaurus.org
%% You will be asked to selected these concepts during the submission process
%% but this old "keyword" functionality is maintained in case authors want
%% to include these concepts in their preprints.

%% From the front matter, we move on to the body of the paper.
%% Sections are demarcated by \section and \subsection, respectively.
%% Observe the use of the LaTeX \label
%% command after the \subsection to give a symbolic KEY to the
%% subsection for cross-referencing in a \ref command.
%% You can use LaTeX's \ref and \label commands to keep track of
%% cross-references to sections, equations, tables, and figures.
%% That way, if you change the order of any elements, LaTeX will
%% automatically renumber them.
%%
%% We recommend that authors also use the natbib \citep
%% and \citet commands to identify citations.  The citations are
%% tied to the reference list via symbolic KEYs. The KEY corresponds
%% to the KEY in the \bibitem in the reference list below. 

\section{Introduction} \label{sec:intro}

Measurements of cometary compositions provide a unique tool for investigating ice chemistry in the protosolar disk midplane during the epoch of planet formation, and can therefore provide insight into the reagents available for pre-biotic chemistry in the early Solar System \citep{mum11}. A wealth of molecular species were recently detected in comet 67P by the Rosetta spacecraft \citep{alt19}, yet remote observations of coma gases remain the most common method for determining cometary compositions \citep{coc15}.  

The coma is typically understood in terms of a quasi-spherical expanding outflow of `parent' species, sublimating directly from the nucleus, with `daughter' species originating from photolysis of the parents in the coma \citep{has57}. Several coma molecules, on the other hand, exhibit `distributed' (spatially extended)  sources, some of which are believed to arise from the breakdown of macromolecular or dust-grain precursors \citep{mei93,cot08,cor14,cor17a}, the precise identity of which remains unknown, but could be related to the organic-rich, refractory material identified in comet 67P by the Rosetta mission \citep{cap15,fra16,bar17}. An analysis of a sample of 26 comets led \citet{mum17} to propose that thermal dissociation of ammonium salts (NH$_4$$^+$X$^-$, where X$^-$ is a deprotonated acid) could be another source of gas-phase coma molecules, and this was found to be a plausible explanation for the abundances of several organics observed in the coma of 67P (\citealt{alt20}; \citealt{poc20}).

To-date, interferometric observations using the Atacama Large Millimeter/submillimeter Array (ALMA) have confirmed the presence of daughter (or distributed) sources of H$_2$CO in three comets, while the HNC and CS molecules have been found to exhibit distributed sources in two comets \citep{cor14,rot21,biv22}. Earlier single-dish mapping work \citep{biv99} and interferometric observations \citep{mil06} identified extended H$_2$CO spatial distributions in comets C/1996 B2 (Hyakutake) and C/1995 O1 (Hale-Bopp), respectively.  Infrared spectroscopic studies have also indicated the presence of both nucleus and coma sources for H$_2$CO and CO \citep{dis99,dis06}, although no compelling evidence for a distributed CO source was found by \citet{boc10}. Despite decades of investigations into the chemical compositions of cometary comae and nuclei, it is surprising that the chemical origins of these commonly-detected (and relatively simple) coma gases still remain be conclusively determined.  

A common way to parameterize molecular production in comets as a function of distance from the nucleus is using the Haser daughter formula \citep{has57}. Through radiative transfer modeling of coma mapping observations, it is then possible to derive the characteristic distance scale at which a given species is produced (commonly expressed as a production scale length, or parent scale length, $L_p$; \citealt{biv99,cor14,rot21b}).  Due to the paucity of detailed studies to-date, and relatively large uncertainties on the derived $L_p$ values for HNC, H$_2$CO and CS, combined with strong variability in the H$_2$CO parent scale lengths derived for different comets (even after correcting for heliocentric distance), further studies are needed to help improve our understanding of the distributions of these molecules in cometary comae. By comparison with laboratory measurements and detailed numerical models \citep[\emph{e.g.}][]{mei93,fra06,cor21}, observational characterization of molecular production as a function of distance from the nucleus allows proposed identities of the parent species to be tested and validated. Measurements of the parent scale lengths of coma daughter species are therefore important in our quest to better understand the native chemical constituents of comets, from which we gain new insights into the chemical processes that occurred during the earliest history of the Solar System.

Previous interferometric studies of molecular production scales in comets have been restricted to bright, long-period comets from the Oort cloud. But these objects represent only part of the wider population of small icy bodies available for study in our Solar System today.  Gravitational scattering by the giant planets is believed to have redistributed comets {within} the inner Solar System and into their various present-day reservoirs: in particular, the Oort cloud and the Kuiper Belt or scattered disk. Most comets discovered each year come from the Oort cloud, while the scattered Kuiper disk is considered to be the source of Jupiter-family comets (JFCs). However, dynamical models have various predictions regarding the formative regions of comets, spanning a diverse range of heliocentric distances (\emph{e.g.} \citealt{gom05,tsi05}). It is therefore vitally important to study both Oort cloud comets (OCCs) and JFCs because it is unclear whether comets from each reservoir were formed in entirely overlapping regions in the protosolar disk. Comparative studies of JFCs and OCCs can also provide insights into the impact of Solar irradiation and thermal processing on the nucleus composition, since JFCs are typically subject to repeated cycles of irradiation, heating and cooling during their relatively frequent, periodic perihelion passages \citep{mee04,gko22}. Due to its extremely close approach to Earth ($\Delta=0.077$~au on UT 2018-12-16), the 2018 apparition of comet 46P/Wirtanen provided a unique opportunity to observe the molecular coma of a Jupiter-family comet, at unprecedented spatial resolution from the ground.

In this study, we present ALMA observations of molecular emission from comet 46P/Wirtanen, conducted in early December 2018 just before its closest approach to Earth. The resulting spectral images include molecules previously identified as daughter/distributed species in Oort cloud comets (H$_2$CO, HNC and CS), as well as suspected parent species (HCN and CH$_3$OH). We also present the first map of CH$_3$CN, which is yet to be imaged in any comet, so its association with sublimating ices in the nucleus, or chemistry in the coma, remains unexplored. Through application of our recently-developed, 3D radiative transfer and excitation model \citep[SUBLIME;][]{cor22}, the coma temperature distribution is derived, parent scale-lengths and abundances are calculated, leading to new insights into the nucleus \emph{vs.} coma contributions of the observed molecules.

\section{Observations}

ALMA observations of 46P/Wirtanen were carried out on UT 2018-12-02 and 2018-12-07, when the comet was $\approx0.1$~au from Earth ($1.1$~au from the Sun), using forty-three 12~m antennas in an intermediate array configuration (with baselines in the range 15-952~m; see Table \ref{tab:obs}). The comet was tracked, and the position of the array phase center on the sky was updated in real-time using JPL Horizons orbital solution \#K181/6. Weather conditions were very good throughout, with a vertical precipitable water vapor column (PWV) of less than 1~mm. This resulted in good phase stability, which was checked and corrected for with regular visits (every 6--7 minutes) to the nearby phase-calibration quasar J0241-0815.

The ALMA correlator was configured to observe three spectral setups in receiver bands 6 and 7, covering lines from HCN (setting 1), HNC and H$_2$CO (setting 2), and CH$_3$OH, CH$_3$CN and CS (setting 3), with spectral resolutions in the range 122--977~kHz (0.1--1.1~km\,s$^{-1}$). Integration times on source and other observational parameters are given in Table \ref{tab:obs}, while the detected spectral line frequencies of interest to the present study are shown in Table \ref{tab:lines}.

{Noisy outlier data points were identified and flagged (removed) through inspection by Joint ALMA Observatory (JAO) staff}. The raw data (visibilities) were further flagged and calibrated using the CASA software (version 5.4; \citealt{casa22}), using standard scripts supplied by the JAO. Prior to imaging, the visibilities were continuum-subtracted using a 2nd-order polynomial fit to the line-free channels in each spectral window. The time-resolved interferometric data series was Doppler-corrected to the rest frame of the comet using the CASA {\tt cvel} task, with cubic spline interpolation between the frequency channels.

Imaging was performed using the CASA {\tt tclean} (Clark) algorithm with natural weighting. A pixel size of $0.1''$ was used for  the band 6 data and $0.05''$ for band 7. Deconvolution of the spatial point spread function (PSF) was performed within an $8''$-diameter circular mask centered on the comet, and stopping at a flux threshold of twice the RMS noise level ($\sigma$). The resulting image cubes were corrected for the response of the ALMA primary beam and then transformed from celestial coordinates to sky-projected distances with respect to the center of the comet, which was determined from the peak of the HCN emission (in settings 1 and 2) and the peak CH$_3$OH emission (in setting 3). The coma gas and continuum (dust plus nucleus) emission peaks were both found to be consistent (within $0.2''$) with the JPL Horizons ephemeris position.

Autocorrelation (total power) spectra were also extracted from the ALMA data following the method of \citet{cor19} and \citet{cor20}. The complete set of autocorrelation scans for all antennas was averaged together to form a single total-power spectrum for each molecule, which was then corrected for atmospheric opacity at the mean elevation angle of the observations, and converted to a flux scale (in Janskys) using the beam size and aperture efficiencies from the ALMA Technical Handbook \citep{rem22}.

\begin{table*}
\centering
\caption{Observational parameters \label{tab:obs}}
{\footnotesize
\hspace*{-27mm}\begin{tabular}{ccccccccccccccc}
\hline\hline
Set.&UT Date&UT Time&Int.$^a$&${r_H}^b$&$\Delta^c$&$\dot{\Delta}^d$&$\nu_R$$^e$&Band$^f$&Baselines$^g$&${\theta_{min}}^h$&${\theta_{PB}}^i$&PWV$^j$&${\phi_{STO}}^k$&PS Ang.$^l$\\
&&&(min)&(au)&(au)&(km\,s$^{-1}$)&(GHz)&&(m)&($''$)&($''$)&(mm)&($^{\circ}$)&($^{\circ}$)\\
\hline
1&2018-12-02&04:16--05:12&43&1.07&0.115&$-7.2$&354.5&7&15.1--952&$0.45\times0.31$&16.4&0.63&44.1&36.3\\
2&2018-12-02&02:30--03:52&63&1.07&0.115&$-7.4$&351.8&7&15.1--952&$0.41\times0.31$&16.5&0.90&44.2&36.2\\
3&2018-12-07&02:34--04:06&63&1.06&0.096&$-5.9$&241.8&6&15.1--784&$0.62\times0.55$&24.0&0.78&42.2&38.5\\
\hline
\end{tabular}
}
\parbox{0.9\textwidth}{\footnotesize 
\vspace*{1mm}
$^a$ On-source observing time.\\
$^b$ Heliocentric distance of the comet.\\
$^c$ Geocentric distance of the comet.\\
$^d$ Comet's mean topocentric radial velocity.\\
$^e$ Representative frequency.\\
$^f$ {ALMA receiver band 6 range is 211--275 GHz; band 7 covers 275--370 GHz.}\\
$^g$ Range of antenna baseline lengths.\\
$^h$ Angular resolution (dimensions of Gaussian fit to PSF) at $\nu_R$.\\
$^i$ Primary beam FWHM at $\nu_R$.\\
$^j$ Median precipitable water vapor column length at zenith.\\
$^k$ Sun-Target-Observer (phase) angle.\\
$^l$ Position angle (in the plane of the sky) of the extended Sun-target vector, counter-clockwise from north.\\
}
\end{table*}

\begin{table}
\centering
\caption{Observed spectral line parameters\label{tab:lines}}
\begin{tabular}{lcccc}
\hline
\hline
Species&Transition&Freq.&$E_{u}$&Res.\\
&&(MHz)&(K)&(kHz)\\
\hline
   CH$_3$OH & $5_0$--$4_0$\,$E$ & 241700.1590(0040) & 48 & 244\\
   CH$_3$OH & $5_{-1}$--$4_{-1}$\,$E$ & 241767.2340(0040) & 40 & 244\\
   CH$_3$OH & $5_0$--$4_0$\,$A^+$ & 241791.3520(0040) & 35 & 244\\
   CH$_3$OH & $5_4$--$4_4$\,$A^{\pm}$ & 241806.5240(0040) & 115 & 244\\
   CH$_3$OH & $5_{-4}$--$4_{-4}$\,$E$ & 241813.2550(0040) & 123 & 244\\
   CH$_3$OH & $5_3$--$4_3$\,$A^{\pm}$ & 241833.1060(0040) & 85 & 244\\
   CH$_3$OH & $5_2$--$4_2$\,$A^-$ & 241842.2840(0040) & 73 & 244\\
   CH$_3$OH & $5_3$--$4_3$\,$E$ & 241843.6040(0040) & 83 & 244\\
   CH$_3$OH & $5_{-3}$--$4_{-3}$\,$E$ & 241852.2990(0040) & 98 & 244\\
   CH$_3$OH & $5_1$--$4_1$\,$E$ & 241879.0250(0040) & 56 & 244\\
   CH$_3$OH & $5_2$--$4_2$\,$A^+$ & 241887.6740(0040) & 73 & 244\\
   CH$_3$OH & $5_{-2}$--$4_{-2}$\,$E$ & 241904.1470(0040) & 61 & 244\\
   CH$_3$OH & $5_2$--$4_2$\,$E$ & 241904.6430(0040) & 57 & 244\\
         CS & $5$--$4$ & 244935.5565(0028) & 35 & 488\\
   CH$_3$CN & $14_4$--$13_4$ & 257448.1282(0002) & 207 & 977\\
   CH$_3$CN & $14_3$--$13_{-3}$ & 257482.7919(0002) & 157 & 977\\
   CH$_3$CN & $14_{-3}$--$13_3$ & 257482.7919(0002) & 157 & 977\\
   CH$_3$CN & $14_2$--$13_2$ & 257507.5619(0002) & 121 & 977\\
   CH$_3$CN & $14_1$--$13_1$ & 257522.4279(0002) & 100 & 977\\
   CH$_3$CN & $14_0$--$13_0$ & 257527.3839(0002) & 93 & 977\\
    H$_2$CO & $5_{1,5}$--$4_{1,4}$ & 351768.6450(0300) & 62 & 244\\
     HCN & $4$--$3$ & 354505.4773(0005)& 43 & 122\\
     HNC & $4$--$3$ & 362630.3030(0090) & 44 & 244\\
\hline
\end{tabular}
\parbox{0.6\textwidth}{\footnotesize
\vspace*{1mm} Note --- All spectral line data were obtained from the Cologne Database for Molecular Spectroscopy (CDMS) \citep{end16}. Transition quantum numbers are $J'_{K'}$--$J''_{K''}$ for CH$_3$OH and CH$_3$CN, $J'_{K_a',K_c'}$--$J'_{K_a',K_c'}$  for H$_2$CO and $J'$--$J''$ for CS, HCN and HNC. Uncertainties on the trailing digits of the spectral line frequencies are given in parentheses.}
\end{table}

\section{Results}

\begin{figure}
\centering
\includegraphics[height=4.8cm]{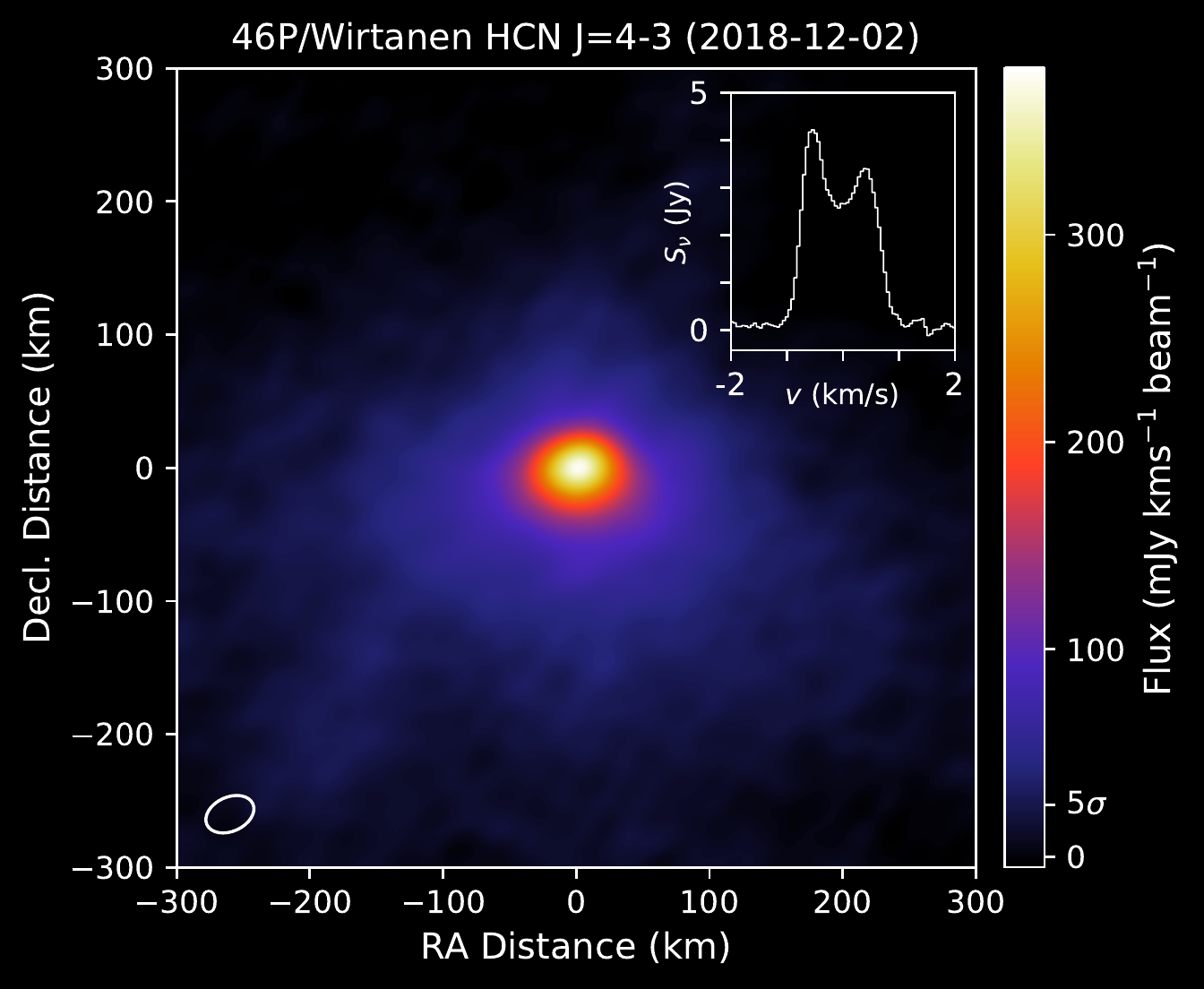} 
\includegraphics[height=4.8cm]{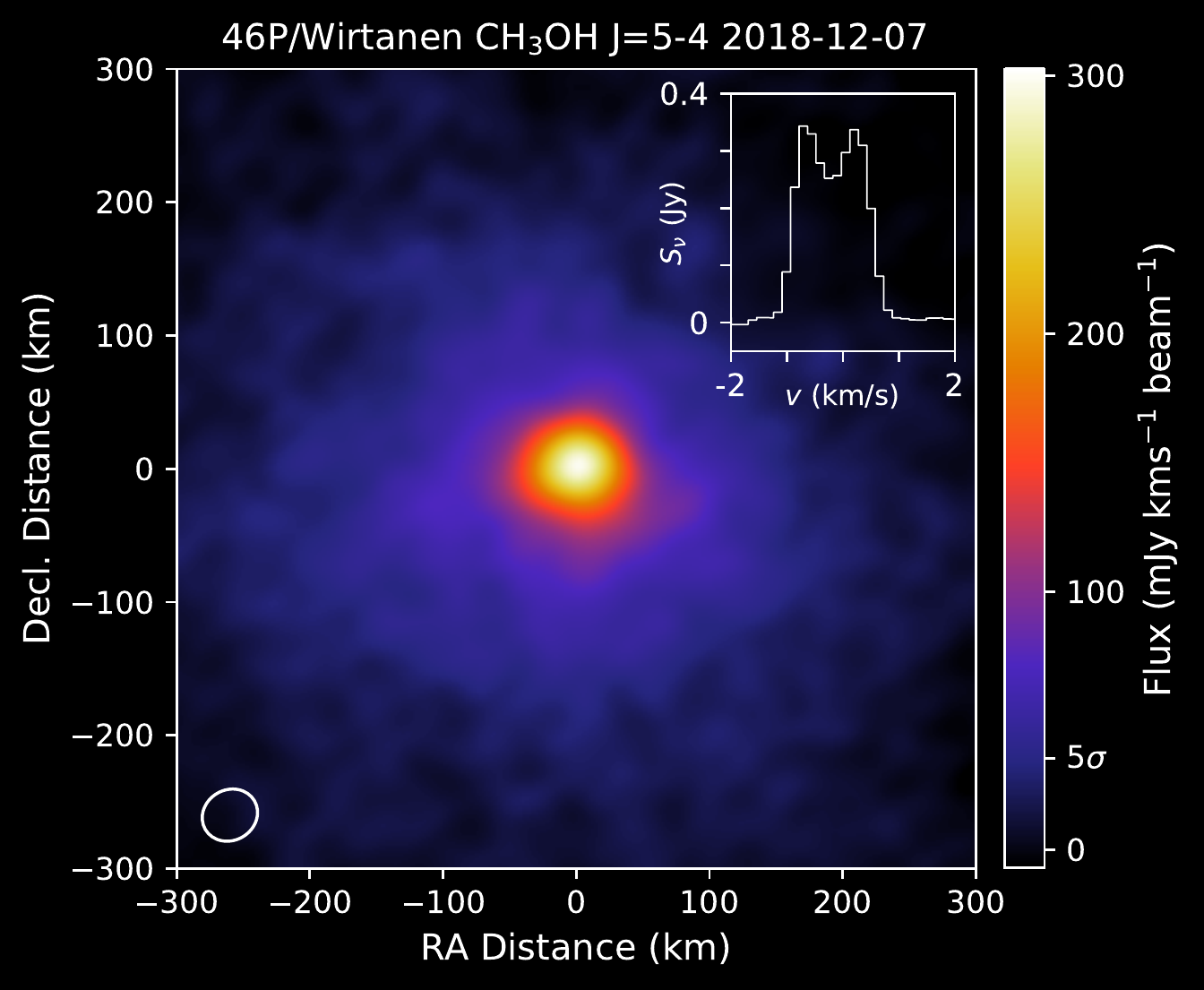}
\includegraphics[height=4.8cm]{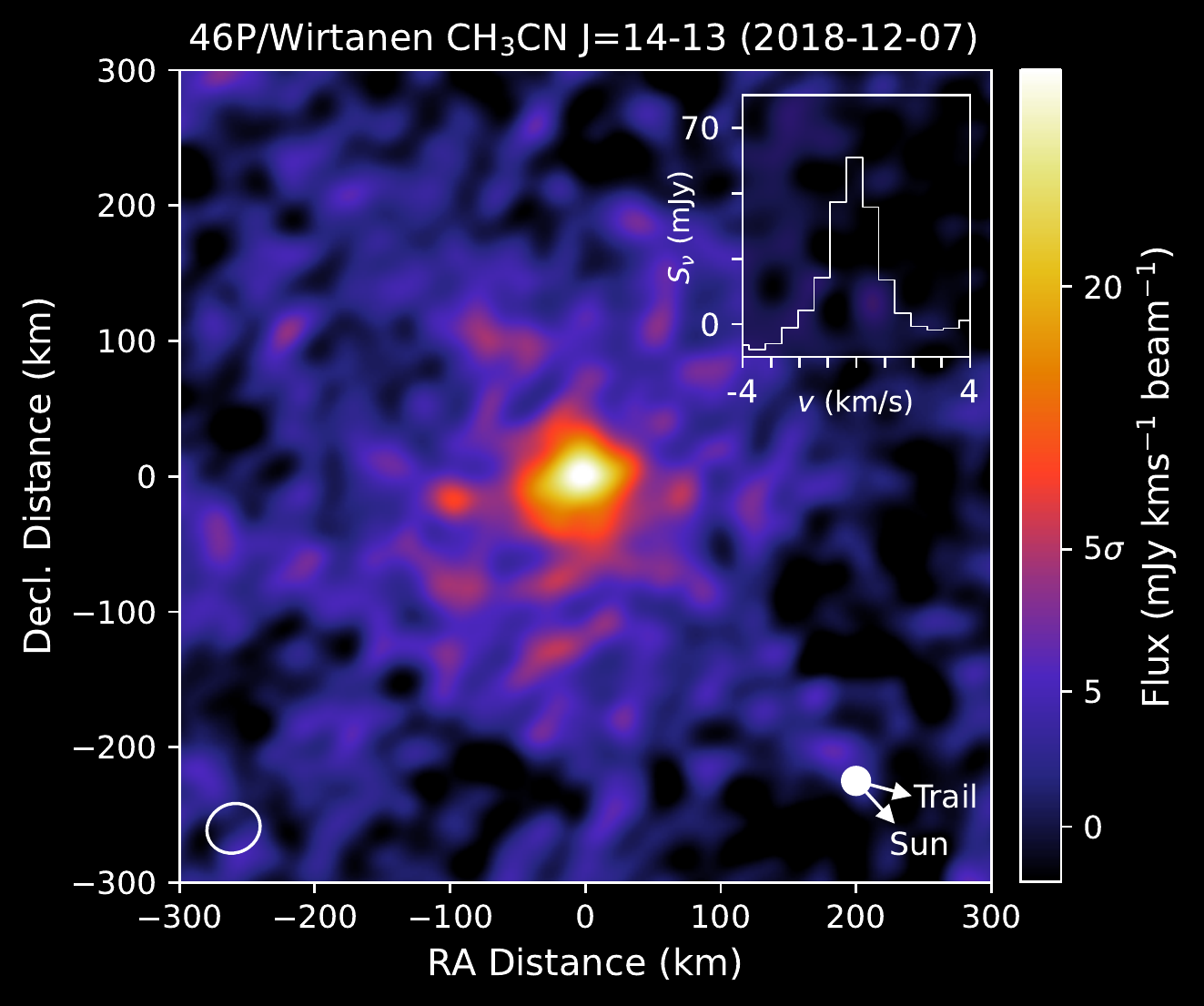}
\caption{Spectrally-integrated ALMA emission maps of HCN, CH$_3$OH and CH$_3$CN, centered on their respective emission peaks. Beam size (FWHM of a Gaussian fit to the PSF) is shown lower left.  Inset plots show the molecular spectra on a cometocentric velocity scale, integrated within a $5''$-diameter aperture centered on the brightness peak. The (sky-projected) direction of the comet's orbital trail and comet-Sun vector are shown for CH$_3$CN in the lower right. Five times the RMS noise level ($5\sigma$) is indicated on the color bar for each species ($\sigma({\rm HCN})=5.0$~mJy\,km\,s$^{-1}$, $\sigma({\rm CH_3OH})=7.1$~mJy\,km\,s$^{-1}$, $\sigma({\rm CH_3CN})=2.1$~mJy\,km\,s$^{-1}$).  \label{fig:parentMaps}}
\end{figure}

\begin{figure}
\centering
\includegraphics[height=4.8cm]{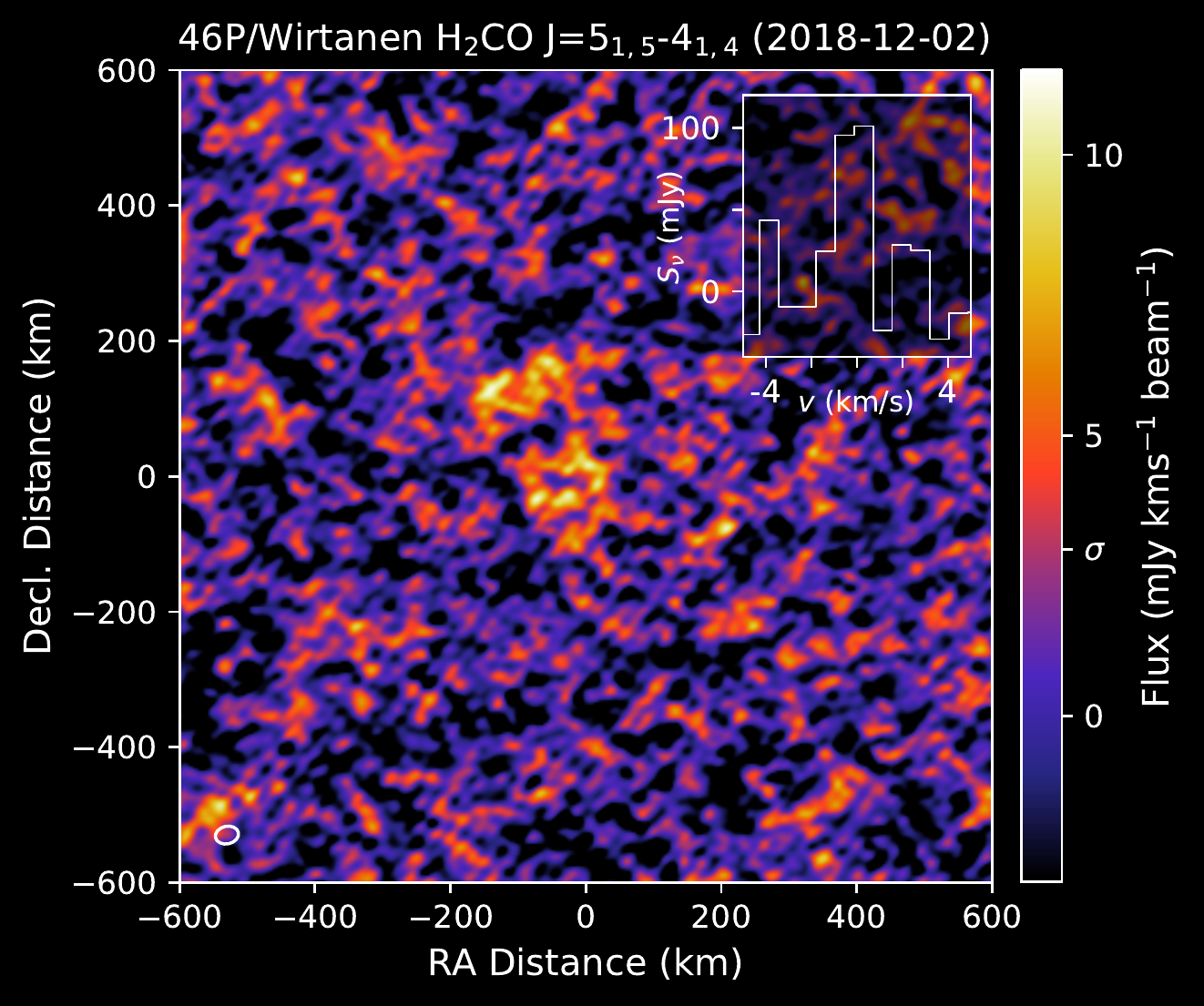} 
\includegraphics[height=4.8cm]{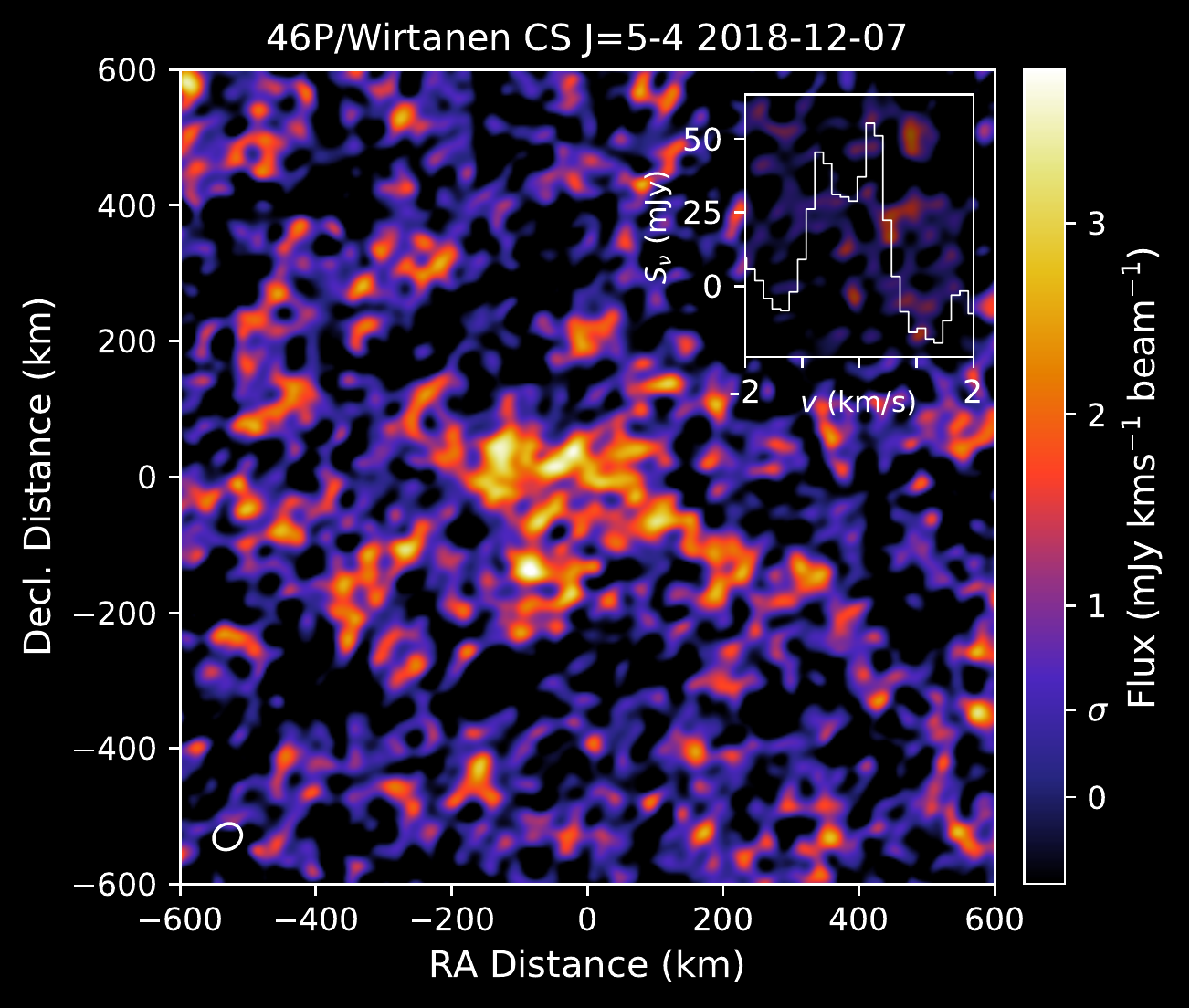}
\includegraphics[height=4.8cm]{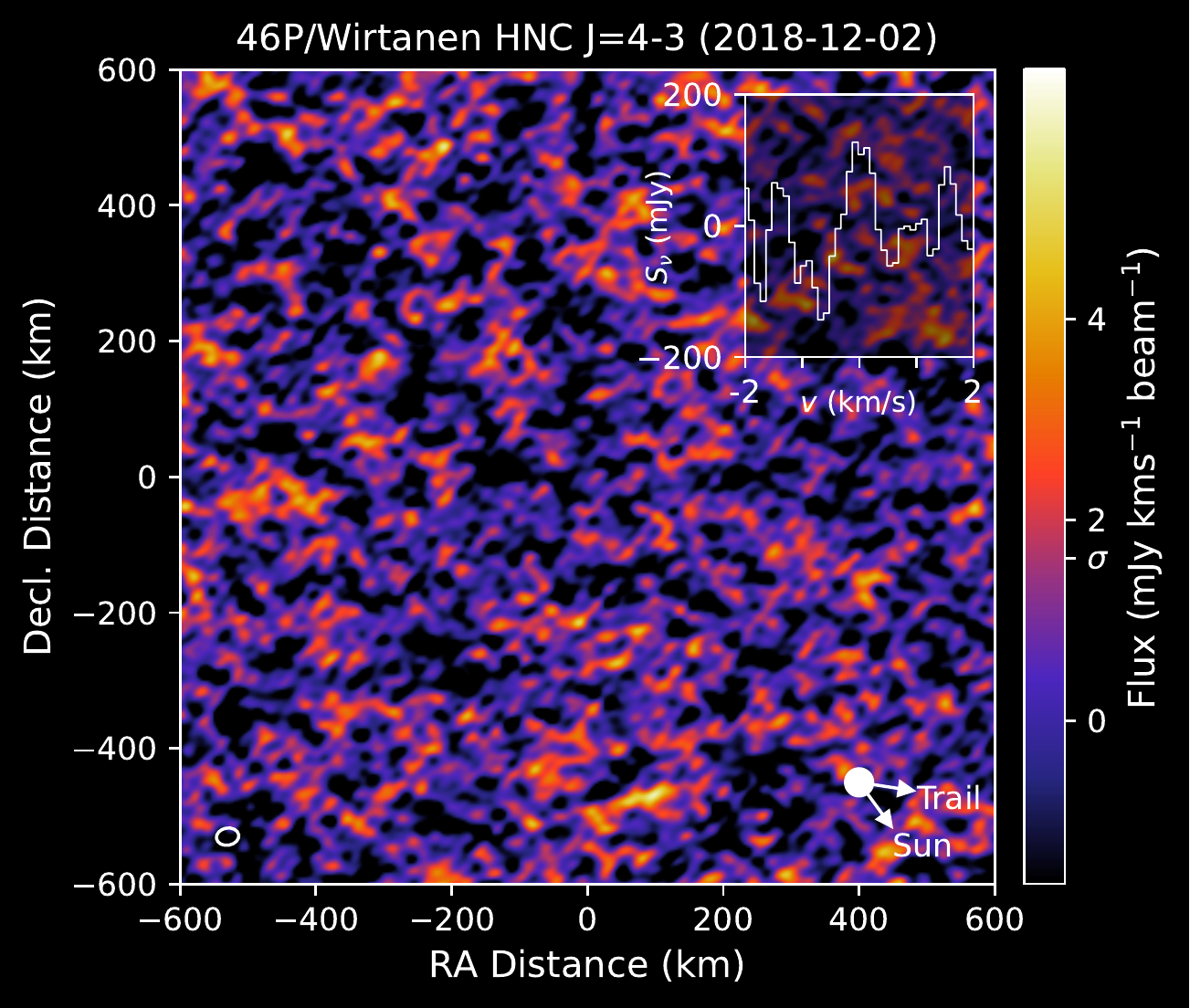}
\caption{Same as Figure \ref{fig:parentMaps}, but for H$_2$CO, CS and HNC. The H$_2$CO and HNC maps are centered on the position of the HCN emission peak, relative to the phase center, while the CS map is centered on the CH$_3$OH peak. In the inset spectrum, the H$_2$CO data have been binned over 8 spectral channels for display. The RMS noise level ($\sigma$) is indicated on the color bar for each species ($\sigma({\rm H_2CO})=3.0$~mJy\,km\,s$^{-1}$, $\sigma({\rm CS})=0.5$~mJy\,km\,s$^{-1}$, $\sigma({\rm HNC})=1.6$~mJy\,km\,s$^{-1}$) \label{fig:daughterMaps}}
\end{figure}

Spectral line emission was identified with at least $3\sigma$ confidence for all the transitions listed in Table \ref{tab:lines}. For these lines, the spectral channels with fluxes above $2\sigma$ were integrated to produce the flux maps shown in Figures \ref{fig:parentMaps} and \ref{fig:daughterMaps}. A centrally-peaked morphology is clearly evident for HCN, CH$_3$OH and CH$_3$CN (Figure \ref{fig:parentMaps}), whereas the H$_2$CO and CS fluxes (Figure \ref{fig:daughterMaps}) are more spatially distributed, lacking any prominent, well-defined emission peak. HNC does not show any significant emission in the interferometric data.

\begin{figure}
\centering
\includegraphics[height=4cm]{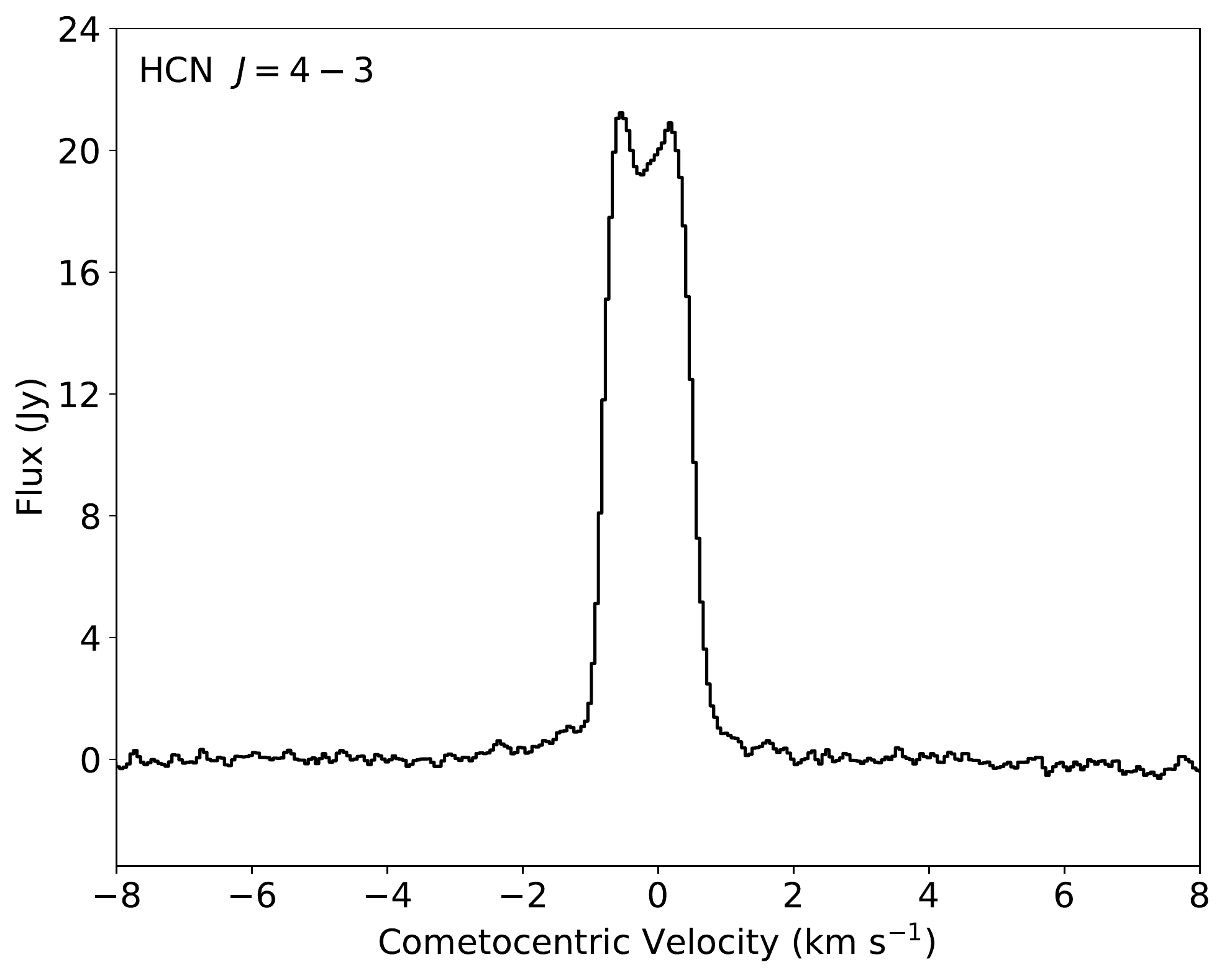}
\includegraphics[height=4cm]{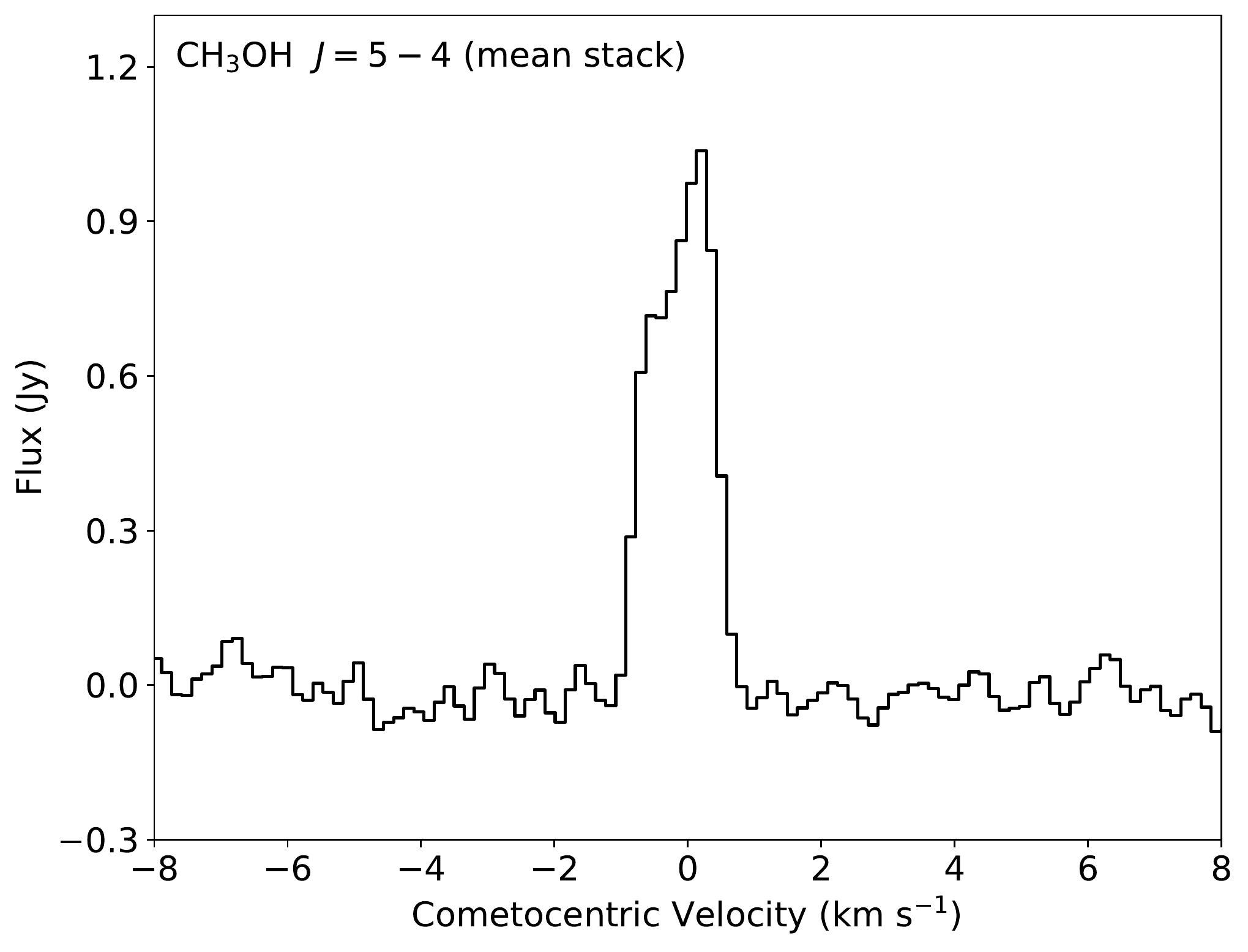}
\includegraphics[height=4cm]{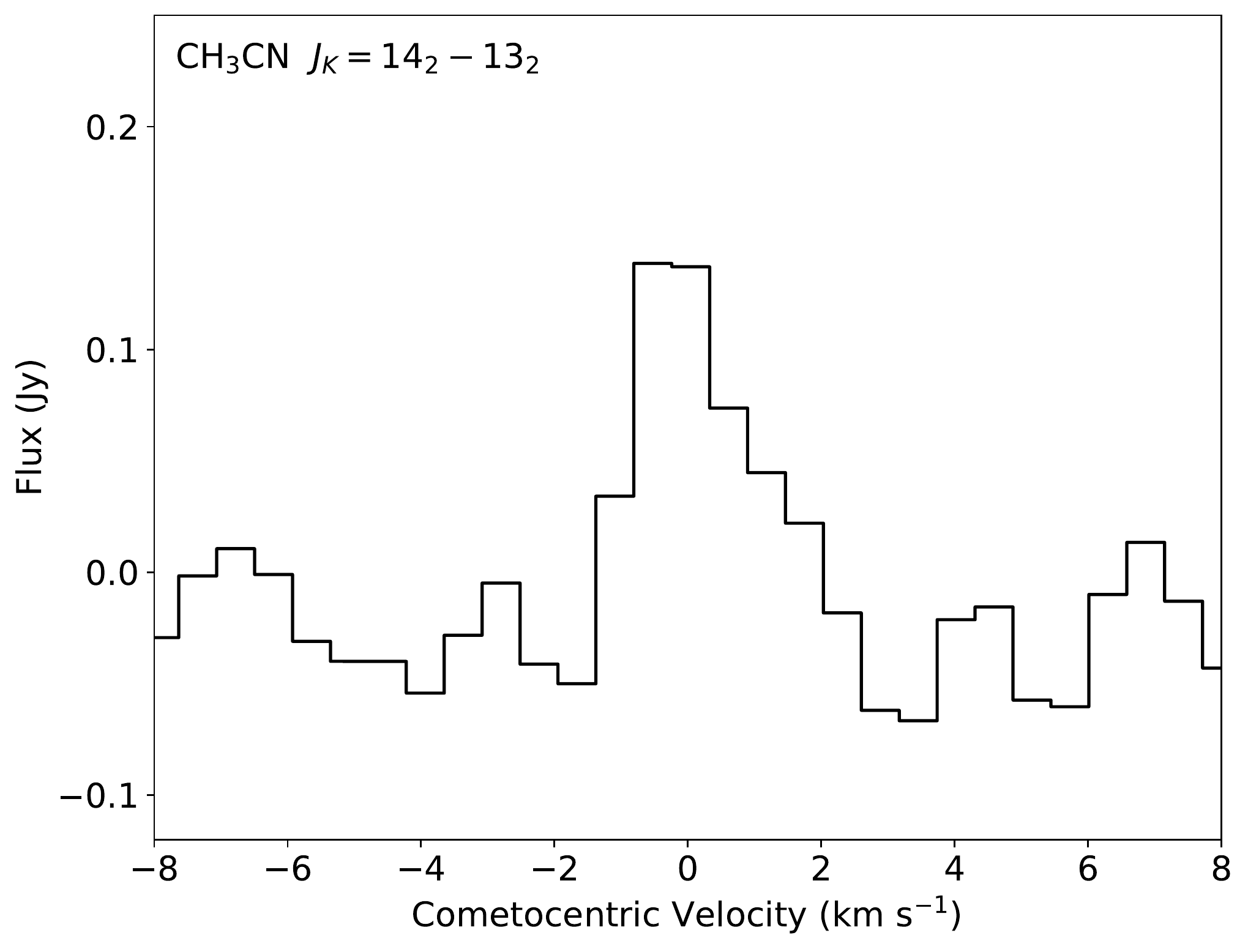}
\includegraphics[height=4cm]{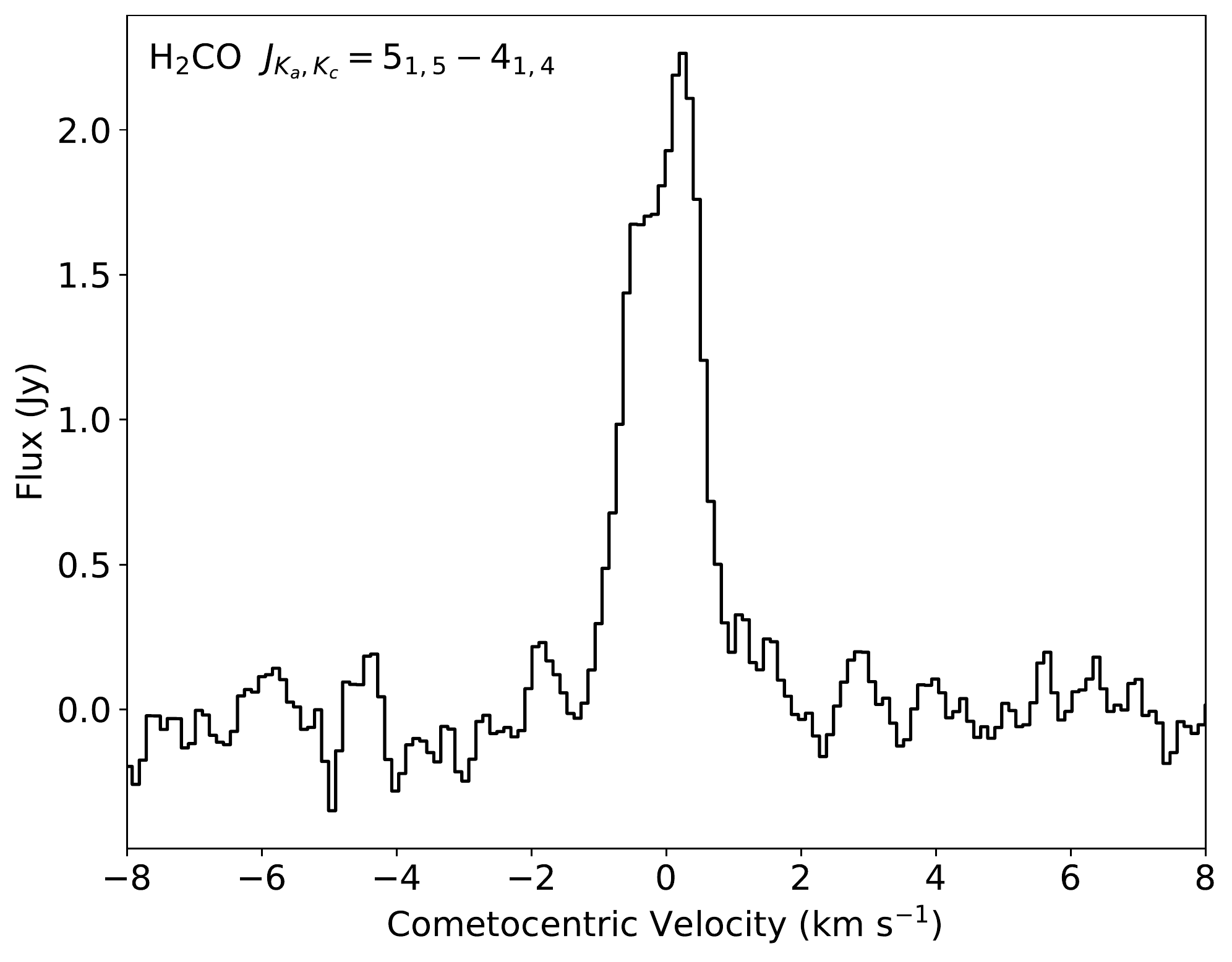}
\includegraphics[height=4cm]{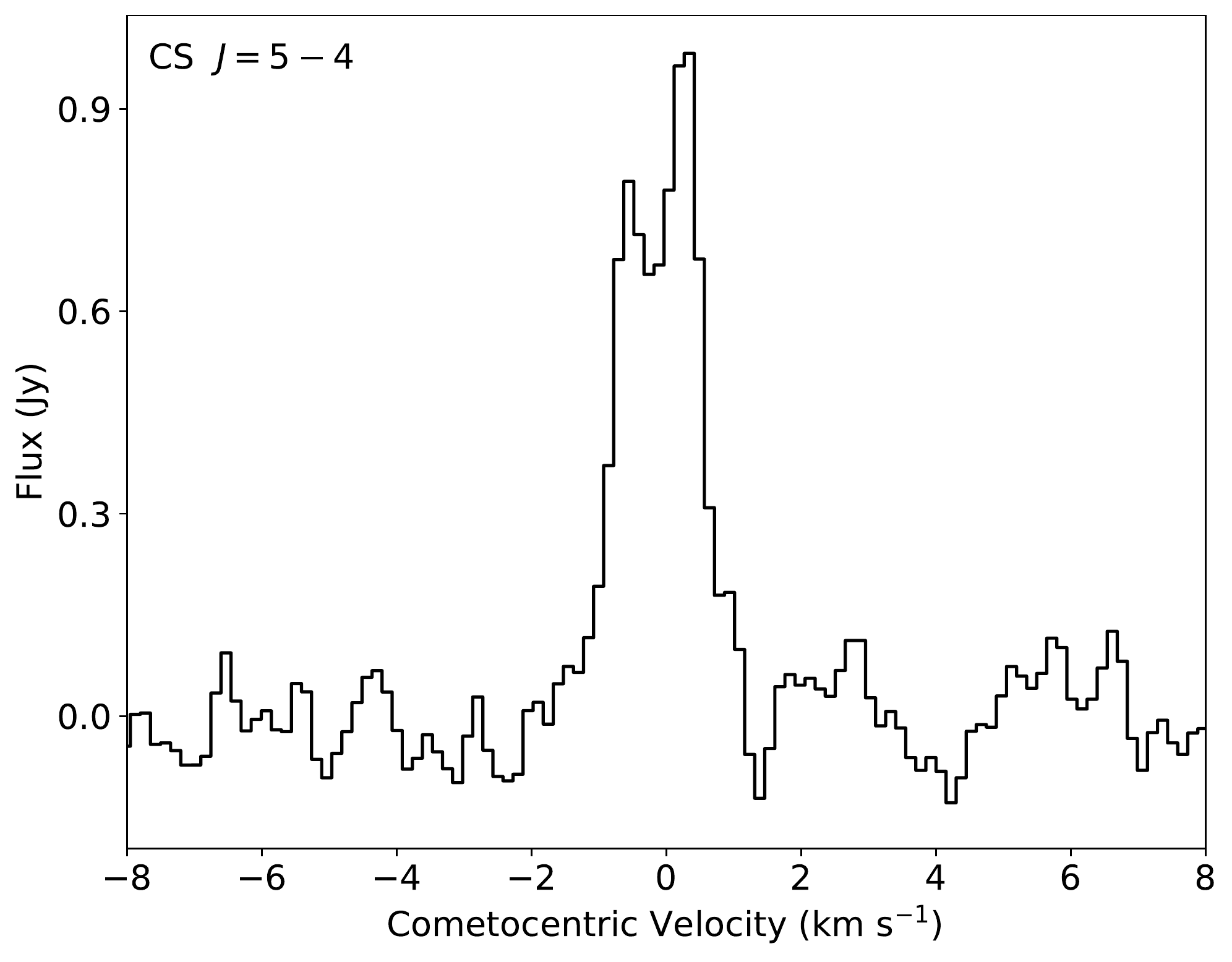}
\includegraphics[height=4cm]{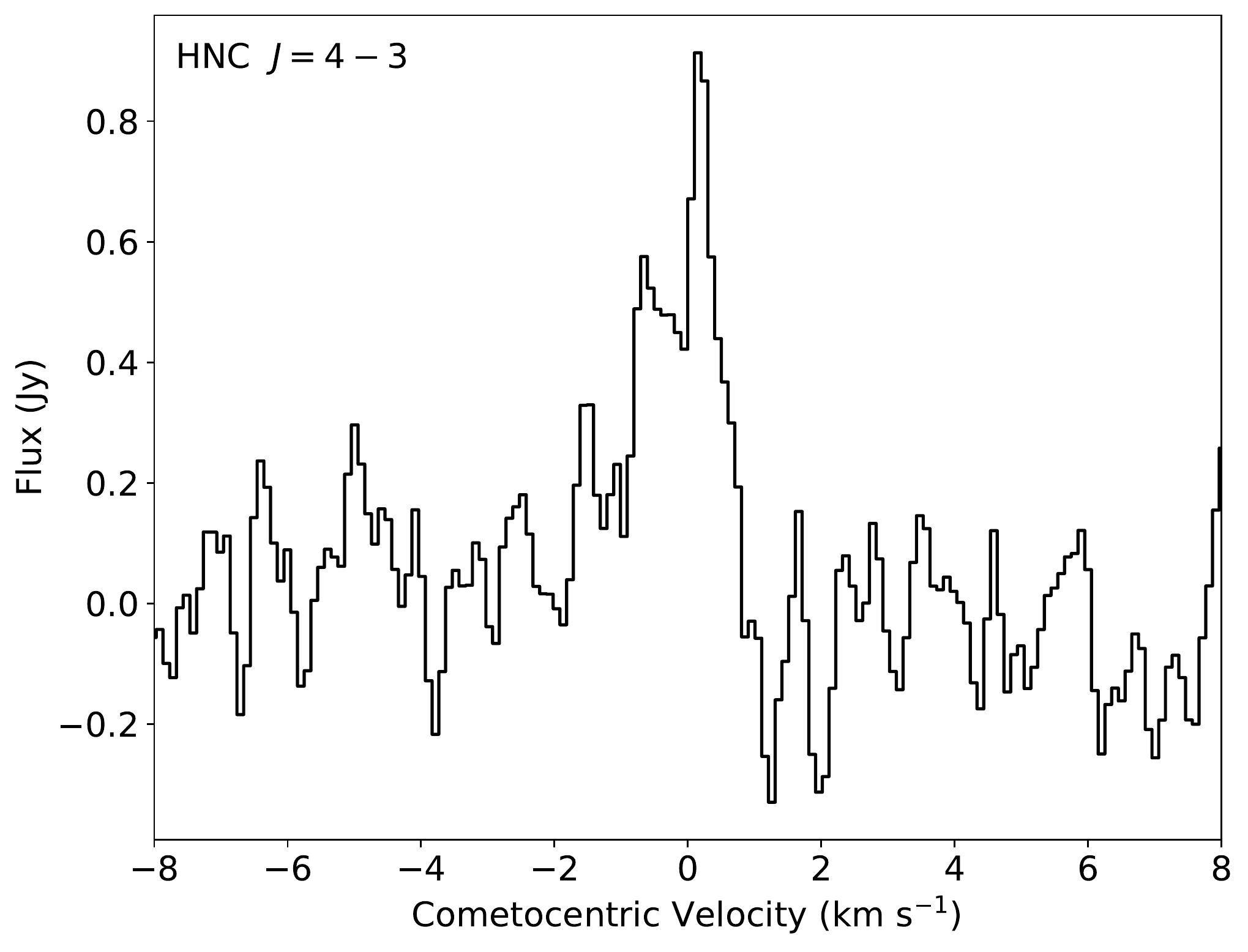}
\caption{ALMA autocorrelation (total power) spectra, in the rest frame of the comet. The spectra have been baseline-subtracted and corrected for the nominal ALMA beam efficiency (see text). CH$_3$OH is the mean of the five strongest unblended lines. \label{fig:acorr}}
\end{figure}

The inset panels in the upper-right corners of Figures \ref{fig:parentMaps} and \ref{fig:daughterMaps} show the spectral line fluxes of each species as a function of cometocentric Doppler velocity, integrated within a $5''$-diameter circle centered on the comet. For CH$_3$OH, the inset spectrum represents an average of the five strongest, spectroscopically distinct (unblended) lines, whereas for CH$_3$CN, the four strongest lines were averaged. The autocorrelation spectra for each species are shown in Figure \ref{fig:acorr}. For HNC, emission is clearly detected in the autocorrelation spectrum, but not in the interferometric data.

In contrast to the interferometric data, which have a spatial resolution $\sim0.3$--$0.6''$ (see Table \ref{tab:obs}), or 26--38~km at the distance of the comet, and are insensitive to any structures on the sky larger than $\sim3$--$6''$ (260--410~km), the autocorrelation spectra contain flux from the entirety of the ALMA primary beam (which has a FWHM $=16$--$24''$). Due to the extended nature of the cometary coma (spanning hundreds of arc-seconds), the autocorrelation spectra contain intrinsically more flux per beam, especially for gases that have increasing abundances as a function of distance from the nucleus.  The weaker, more spatially extended appearance of the H$_2$CO, CS and HNC emission maps, combined with the relative strength of their autocorrelation spectra, is therefore characteristic of a more extended spatial distribution for these species, consistent with their release in the coma as daughter/distributed species. On the other hand, the strongly centrally peaked morphologies for HCN, CH$_3$OH and CH$_3$CN are indicative of their production as parent species, directly from the nucleus. However, considering the observed flux distributions result from a complex interplay between molecular excitation and emission processes in a three-dimensional outflowing coma, detailed radiative transfer modeling is required to reliably determine the molecular origins and derive production scale lengths.

\section{SUBLIME Radiative Transfer Modeling}

\subsection{Model physics and geometry}

Our modeling approach follows a similar method to the previous comet interferometry studies by \citet{boi07}, \citet{cor14} and \citet{rot21}. Here, the interferometric visibilities are modeled using a non-LTE radiative transfer code called SUBLIME (SUBlimating gases in LIME, where LIME is the LIne Modeling Engine; \citealt{bri10}), which takes into account the detailed molecular excitation and emission processes that occur in the presence of varying coma density, temperature and abundance distributions. The model incorporates a \citet{has57} radial density profile, and treats coma molecules as parent species, photochemical daughter species, or a mixture of both, expanding outward at a constant velocity. However, our method differs from the 1D modeling performed by earlier studies as we adopt a two-component description of the outflowing gases in three dimensions, similar to that used for the analysis of single-dish CO spectral/spatial observations of comet C/2016 R2 (PanSTARRS) by \citet{cor22}.

The coma is divided into two solid-angle regions ($\Omega_1$, $\Omega_2$), each with an independent water production rate ($Q_1$, $Q_2$), outflow velocity ($v_1$, $v_2$) and kinetic temperature as a function of radius ($T_1(r)$, $T_2(r)$). As shown in Appendix \ref{sec:geom} Figure \ref{fig:geom}, the first solid angle region ($\Omega_1$) is defined by a cone of half-opening angle $\theta_{jet}$, with its apex at the center of the nucleus, and its axial vector pointing at a phase angle $\phi$ with respect to the observer, and at a position angle $\psi$ in the plane of the sky. The second region ($\Omega_2$) represents the remaining (ambient) coma. This geometry, although still likely to be highly simplified compared to reality, represents an evolution from the two-hemisphere coma model of \citet{rot21b}, which was successfully used for the analysis of asymmetric CH$_3$OH spectra observed in 46P using the ALMA 7~m array (ACA). As explained by \citet{cor22}, the SUBLIME model provides a sufficiently good approximation to the structure of a coma with a single dominant, rapidly expanding (near-) sunward-facing jet, embedded in a slower-moving, ambient coma. The adopted geometry is found to reproduce very well the spectral line profiles (including asymmetries) observed for all our detected species, while keeping the number of variable model parameters at a minimum.

The radial distribution of a daughter species is governed by its parent scale length ($L_p$), which, in a uniformly-expanding coma of outflow velocity $v$, is related to the photodissociation rate ($\Gamma$) of its parent species by $L_p=v/\Gamma$. For consistency, we assume that $\Gamma$ is constant for a given parent species across both coma solid angle regions ($\Omega_1$, $\Omega_2$), so the ratio of production scale-lengths for a daughter species in those two regions is determined by the ratio of outflow velocities ($v_1/v_2$). Consequently, rather than reporting parent scale lengths for both regions, we report only the $L_p$ value corresponding to the jet component ($\Omega_1$). Outflow velocities for daughter species in our model are allowed to differ from the parents. Daughter production rates in the two coma solid angle regions are also independently optimized; reported daughter abundances are therefore taken as averages over the entire coma.

The excitation calculation in our model is time-dependent and takes into account radiative cooling of rotational levels, (de-)excitation by collisions with H$_2$O and electrons, and pumping by the Solar radiation field as the gas moves outward (for more details see \citealt{cor22}). The impact of coma opacity on the molecular excitation is negligible for the observed species in this comet, so radiation trapping effects are not included.  We adopt H$_2$O--HCN collision rates from \citet{dub19} (also assumed to apply to HNC), while collision rates of H$_2$O with CH$_3$OH, H$_2$CO, CH$_3$CN and CS are assumed to be the same as for H$_2$, and have been taken from the LAMDA database \citep{van20}. An electron density scaling factor of $x_{ne}=0.2$ is used, following the recommendations of \citet{har10} and \citet{biv19,biv21}. Pumping rates for HCN, HNC, H$_2$CO, and CS were calculated as described by \citet{cor19} and \citet{rot21b}, based on the method of \citet{cro83a}, with rovibrational transition data from the latest versions of the HITRAN and GEISA catalogues \citep{gor22,del21}. CH$_3$OH pumping rates were calculated similarly. However, the Einstein $A$ coefficients for CH$_3$OH in the HITRAN catalogue were found to be incorrect because they were calculated considering corrupted statistical weights and partition functions. On the other hand, HITRAN line intensities at 296~K are accurate, so by converting these to $A$ coefficients, employing accurate statistical weights and partition functions \citep{vil12}, the required spectroscopic information could be recovered. We further complemented the HITRAN data by adding rovibrational transitions from the strong $v_2$, $v_3$ and $v_9$ bands from the quantum band models of \citet{vil12}. The data were then homogenized based on the CH$_3$OH quantum numbers and energies reported for the ground-state Hamiltonian by \citet{mek99}.

\subsection{Model optimization strategy}
\label{sec:model}

Our ALMA observations were obtained over two dates: 2018-12-02 and 2018-12-07 (see Table \ref{tab:obs}). On 2018-12-02, the coma physical structure (characterized by the ratio of production rates in the sunward jet \emph{vs.} ambient coma, $Q_1/Q_2$, and the jet properties, $\theta_{jet}$, $\phi$ and $\psi$) was determined using a model fit to the bright HCN $J=4-3$ line observations. On 2018-12-07, multiple lines from the CH$_3$OH ($J_K=5_K-4_K$) band were used instead, since HCN was not observed on this date. Both these species correlate well with H$_2$O \citep{del16,boc17}, and therefore provide a reasonable proxy for the overall coma outflow velocity and $Q_1/Q_2$ ratio in the absence of spectrally-resolved H$_2$O data. The H$_2$O coma physical structure derived from fitting the HCN data was therefore used in our models for H$_2$CO and HNC (also obtained on 2018-12-02), whereas the structure derived using CH$_3$OH was applied to CH$_3$CN and CS. For CH$_3$CN, the observed ALMA data were of insufficient spectral resolution to adequately constrain the values of $v_1$ and $v_2$, whereas for H$_2$CO, HNC and CS, the correlation with H$_2$O is insufficiently demonstrated, and the data were of insufficient S/N to reliably constrain the coma structure by themselves.

The total H$_2$O production rate on each date was obtained from a linear fit to the measurements between 2018-12-02 and 2018-12-10 by \citet{com21} using Ly-$\alpha$ observations by the SOlar and Heliospheric Observatory (SOHO) satellite, which gave $Q({\rm H_2O})=6.1\times10^{27}$~s$^{-1}$ on 2018-12-02 and $Q({\rm H_2O})=7.2\times10^{27}$~s$^{-1}$ on 2018-12-07. These values are consistent with a slow increase in $Q({\rm H_2O})$ as the comet approached perihelion on 2018-12-12, and are in line with the average values of $8\times10^{27}$~s$^{-1}$ measured by \citet{lis19} using the Stratospheric Observatory for Infrared Astronomy (SOFIA) between 2018-12-14 and 2018-12-20, and $7\times10^{27}$~s$^{-1}$ measured using IRTF between 2018-12-06 and 2018-12-21 \citep{kha23}.

Our models were set up on a 3D Delaunay grid containing 10,000 points distributed pseudo-randomly with a density of points proportional to the logarithm of the radial distance ($r$) from the center of the nucleus (excluding the nucleus itself). The model domain therefore consisted of a spherical coma region extending from the surface of the nucleus (assumed to be a sphere of radius 500~m), to an outer boundary at $r=2\times10^5$~km. The outer boundary was chosen to be large enough so that further increasing it had no significant impact on our model results. Raytracing was performed along the line-of-sight vectors through each grid point, on a frequency grid with a uniform channel spacing of 100~m\,s$^{-1}$, which was then interpolated in two (spatial) dimensions onto the image grid, which consisted of $768\times768$ pixels $0.1''$ in size. The innermost $4\times4$ pixel region of the image was further super-sampled using a $30\times30$ point Cartesian grid, to accurately capture the nonlinear behavior of the coma flux on the smallest relevant radial scales. To simulate the response of the ALMA primary beam, each plane of the synthetic image cubes was multiplied by a 2D Gaussian of FWHM $=1.13\lambda/D$, where $\lambda$ is the wavelength and $D$ is the antenna diameter. For each molecule, the frequency axis of the resulting synthetic image cube was convolved to the spectral resolution of the ALMA observations, followed by cubic spline interpolation to the observed frequency grid. This allowed a channel-by-channel comparison of the model with the observations.

Interferometric observations inherently suffer from incomplete spatial sampling, and the resulting, Fourier-transformed images can be subject to artifacts introduced by gridding, interpolation, and numerical deconvolution. To facilitate accurate modeling of the data, and avoid our model fits becoming biased by image artifacts, we chose to perform all model fitting in the Fourier domain. This requires taking the Fourier transform of the simulated coma image cubes, then sampling each spectral channel with the same set of $uv$ points (antenna baseline lengths and orientations) as the observations, which was performed using the {\tt vis\_sample} code \citep{loo18}. To make the problem computationally tractable, the observed visibilities were first averaged along the time-axis to produce a single (complex) visibility point per baseline, per channel. The chi-squared statistic

\begin{equation}
\chi^2=\sum_{i,j}\left(\frac{(\Re(V^{obs}_{i,j})-\Re(V^{mod}_{i,j}))^2}{\sigma_{\Re}^2} + \frac{(\Im(V^{obs}_{i,j})-\Im(V^{mod}_{i,j}))^2}{\sigma_{\Im}^2}\right) + \sum_{i}\frac{((S^{obs}_{i})-(S^{mod}_{i}))^2}{\sigma_{S}^2}
\end{equation}

was minimized using {\tt lmfit} \citep{new16}, by application of the Levenberg-Marquardt algorithm. Summation of the interferometric residuals (the difference between $V^{obs}$ for the observations, and $V^{mod}$ for the model) was performed over the set of baselines $j=N(N-1)$, where $N$ is the number of ALMA antennas (43), and $i$ is the number of spectral channels. Due to the presence of coma asymmetries, both the real ($\Re$) and imaginary ($\Im$) parts of the complex visibilities were included in the $\chi^2$ calculation. The difference between the model total power spectra ($S^{mod}_{i}$) and the observed autocorrelation spectra ($S^{obs}_{i}$), provide additional, strong constraints on the production scale length ($L_p$) of each species due to the larger angular scales probed by these data. During fitting, the residuals were scaled by the respective standard deviations of the real and imaginary visibilities, and of the autocorrelation spectra ($\sigma_{\Re}$, $\sigma_{\Im}$ and $\sigma_{S}$, respectively), which were calculated from line-free data regions adjacent to each spectral line.

To allow for errors in the comet ephemeris coordinates, a positional (RA, dec.) offset of the model origin from the image center was included as a further pair of free parameters in the model fits. Statistical uncertainties on all model parameters were obtained from the diagonal elements of the {\tt lmfit} covariance matrix.

\subsection{Coma temperature structure derived from CH$_3$OH visibilities}
\label{sec:ch3oh}

Model optimization was performed first for CH$_3$OH, because the multiple transitions observed for this species span a broad range of upper-state energy levels (see Table \ref{tab:lines}), which allow its rotational excitation state to be determined, from which the coma kinetic temperature is derived (see \citealt{cor17b} or \citealt{biv21}, for example). Our method differs from previous studies, however, due to our model's ability to interpret the coma temperature structure in three dimensions, by fitting the variations in individual CH$_3$OH spectral line channels ($V_i$, $S_i$) as a function of spatial ($uv$) coordinate.

\begin{figure}
\centering
\includegraphics[width=0.5\columnwidth]{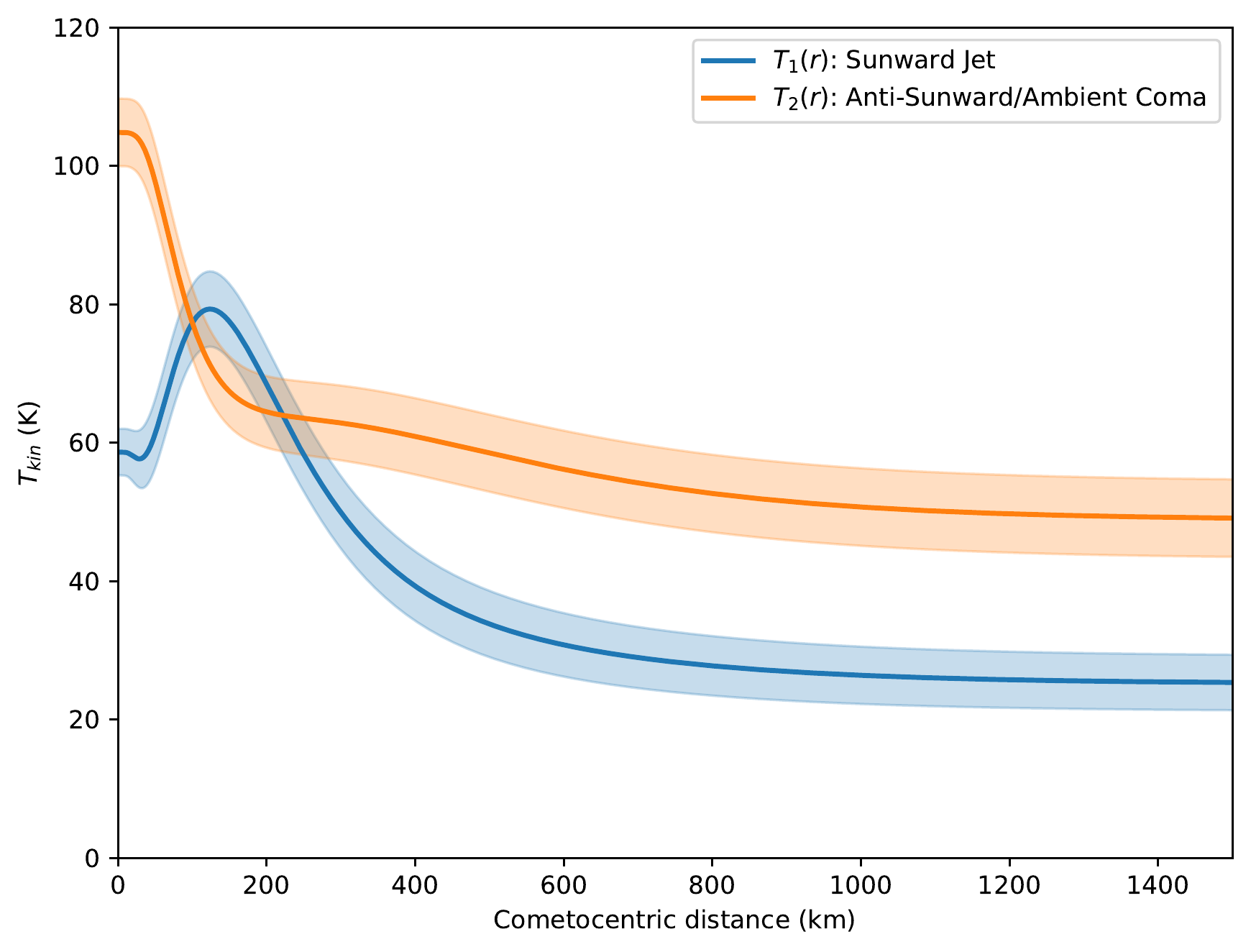}
\caption{Variation of the 46P coma kinetic temperature with radius in the sunward (jet) and antisunward (ambient) coma regions, derived from modeling the CH$_3$OH $J=5-4$ data. Shaded regions indicate the $\pm1\sigma$ error envelopes. \label{fig:Tprof}}
\end{figure}

We began with the simplest assumption of a constant kinetic temperature ($T$) as a function of radius, then added complexity to the $T(r)$ profile until a good fit to the data was obtained. This strategy keeps the number of free parameters at a minimum (therefore keeping the $\chi^2$ minimization computationally feasible), and ensures that there are enough degrees of freedom in the model to reproduce the data, but not so many that the model becomes ill-constrained. To obtain a good fit to the observations, it was necessary to implement different temperature profiles as a function of radius ($T_1(r)$, $T_2(r)$), within the two different coma solid angle regions. Temperature variability was implemented using a segmented linear function (in $\log(r)$--$T$ space, within the domain $r=\theta_{min}$ to $\theta_{PB}$), with a variable number of segments ($n$), of equal length ($l_s$) in $\log(r)$ space. The segmented function was smoothed (in $\log(r)$ space) by convolving it with a Gaussian of FWHM equal to $l_s$. The temperature was set constant inside a radius corresponding to half the minor axis of the ALMA beam ($r=\theta_{min}/2$), and outside a radius corresponding to half the primary beam FWHM ($r=\theta_{PB}/2$). A good fit to the entire CH$_3$OH dataset was obtained using $n=5$ variable points in the temperature profiles as a function of radius (see Figure \ref{fig:Tprof} and Appendix \ref{sec:visfits}, Figure \ref{fig:ch3ohstack}).

Due to the high S/N of our CH$_3$OH observations, the best-fitting coma model is tightly constrained by the data, and has a jet half-opening angle of $\theta_{jet}=70\pm5^{\circ}$, with a phase angle of $\phi=38\pm1^{\circ}$ and position angle of $\psi=33\pm1^{\circ}$. The resulting model jet emanates from close to (within 10$^{\circ}$ of) the sub-solar point on our (assumed) spherical nucleus, and is therefore consistent with preferential outgassing in the general sunward direction. The best-fitting jet outflow velocity is $v_1=0.729\pm0.002$~km\,s$^{-1}$, with the remaining (ambient/nightside) coma expanding at $v_2=0.395\pm0.003$~km\,s$^{-1}$. We initially attempted to fit the observed ALMA data assuming CH$_3$OH was solely a parent species, but a significant ($15\sigma$) improvement in the final $\chi^2$ value was obtained using a composite (parent + daughter) model. Independent optimization of the parent and daughter CH$_3$OH production rates resulted in approximately equal abundances for both components of $1.2\pm0.1$\% (relative to H$_2$O), with a best-fitting parent scale length of $L_p=36\pm7$~km for the CH$_3$OH daughter (in the jet component, $\Omega_1$). The presence of an additional CH$_3$OH daughter component implies the production of significant amounts of CH$_3$OH in the near-nucleus coma, likely from the sublimation of icy grains (see Section \ref{sec:dis_ch3oh}). We also attempted to fit the CH$_3$OH data using a pure daughter model, but this resulted in a statistically worse fit at the $11\sigma$ level, so our modeling strongly implies the presence of both nucleus (parent) and coma (daughter) sources for CH$_3$OH in comet 46P (see also Section \ref{sec:vis}).

A comparison between the best-fitting modeled and observed visibility spectra is shown in Appendix \ref{sec:visfits} Figure \ref{fig:ch3ohstack}, which includes the total power (autocorrelation) spectrum at the top, followed below by the real part of the interferometric visibilities, binned within successive 20~m baseline ranges. The angular scales probed decrease with increasing baseline length, from the $24''$ scale of the autocorrelation spectrum, down to the $0.6''$ scale probed by the longest baselines. We show only the real part of the visibilities since the imaginary components are weak and noisy.

As shown in Figure \ref{fig:Tprof}, the average temperature on the sunward (jet) side of the nucleus is significantly lower than the anti-sunward (ambient) side. This is particularly true closest to the nucleus (within $r=19$~km), where $T_1=59\pm3$~K and $T_2=105\pm5$~K, amounting to almost a factor of two difference. The temperature in the sunward jet rises with increasing cometocentric distance, reaching a peak with $T_1=79\pm5$~K at $r=121$~km (compared with $T_2=71\pm5$~K in the ambient coma at the same radius), before falling smoothly towards $T_1=25\pm4$~K ($T_2=49\pm6$~K) at large radii ($r>1500$~km). The different temperature behavior on opposite sides of the nucleus is remarkable, since it implies significant differences in the balance of heating and cooling mechanisms on the sunward and anti-sunward sides of the comet. \citet{biv21} also identified cooler gas on the day side of the nucleus than the night side (57~K \emph{vs.} 71~K) based on single-dish CH$_3$OH observations probing coma radial distances $\lesssim$600~km. This is qualitatively similar to our result, and was explained by \citet{biv21} as being due to more efficient adiabatic cooling on the sunward side (for further discussion see Section \ref{sec:dis_ch3oh}).

The observed transitions of CH$_3$CN from different $K$ levels provide an additional measure of the coma temperature. However, due to the lower spectral resolution and S/N, it was not possible to reliably constrain the spatial distribution of temperatures for this species. Assuming a constant temperature throughout the coma, we found $T=80\pm8$ K using the CH$_3$CN data.

\subsection{Visibility modeling to derive parent scale lengths}
\label{sec:vis}

Adopting the best-fitting coma kinetic temperature distribution from our CH$_3$OH modeling, we proceeded to optimize the remaining free model parameters for HCN, HNC, CS, H$_2$CO, and CH$_3$CN. As a result of lower S/N and spectral resolution for CS and CH$_3$CN (simultaneously observed with CH$_3$OH), we employed the same jet opening angle ($70^{\circ}$) and ratio of H$_2$O production rates as derived for CH$_3$OH ($Q_1/Q_2=1.21$). The HCN jet opening angle ($70^{\circ}$) and $Q_1/Q_2$ ratio (1.33) was employed for modeling the HNC and H$_2$CO data. The best-fitting HCN coma outflow velocities were $v_1=0.741\pm0.001$~km\,s$^{-1}$ and $v_2=0.443\pm0.001$~km\,s$^{-1}$, with a jet phase angle $\phi=44\pm1^{\circ}$ and position angle $\psi=37\pm1^{\circ}$, consistent with preferential outgassing in the sunward direction. The jet axis was therefore fixed along the comet-sun vector for all species apart from CH$_3$OH.  Model fits to the binned visibility spectra for all molecules are shown in Appendix \ref{sec:visfits} (Figures \ref{fig:ch3ohstack} to \ref{fig:hncstack}). Since the visibility data for CH$_3$CN, H$_2$CO, CS and HNC are noisier than for CH$_3$OH and HCN (particularly on large baselines), some of the longer baseline ranges for which no signal is evident have been omitted from Figures \ref{fig:ch3cnstack} to \ref{fig:hncstack}.

\begin{figure}
\centering
\includegraphics[height=4cm]{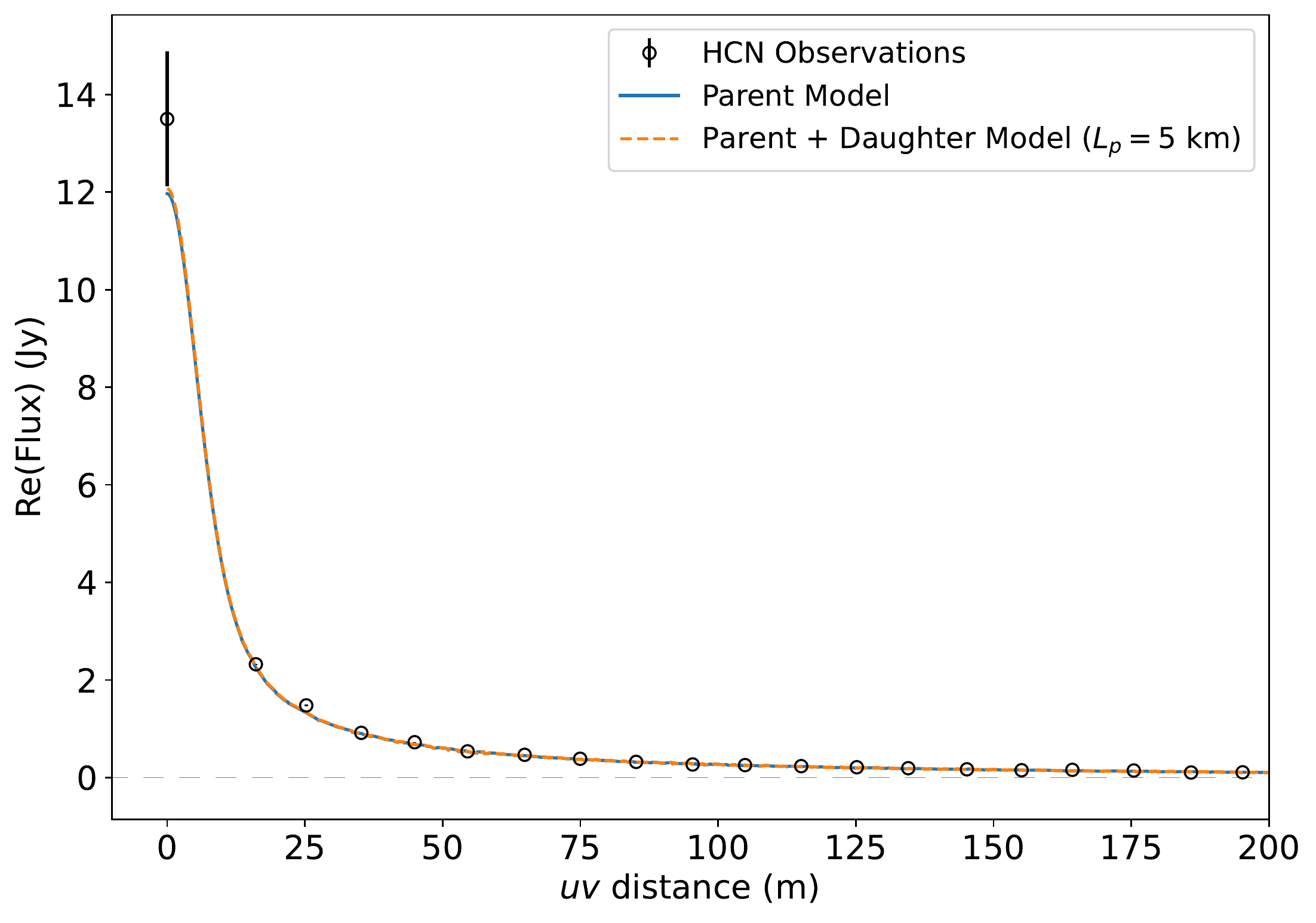}
\includegraphics[height=4cm]{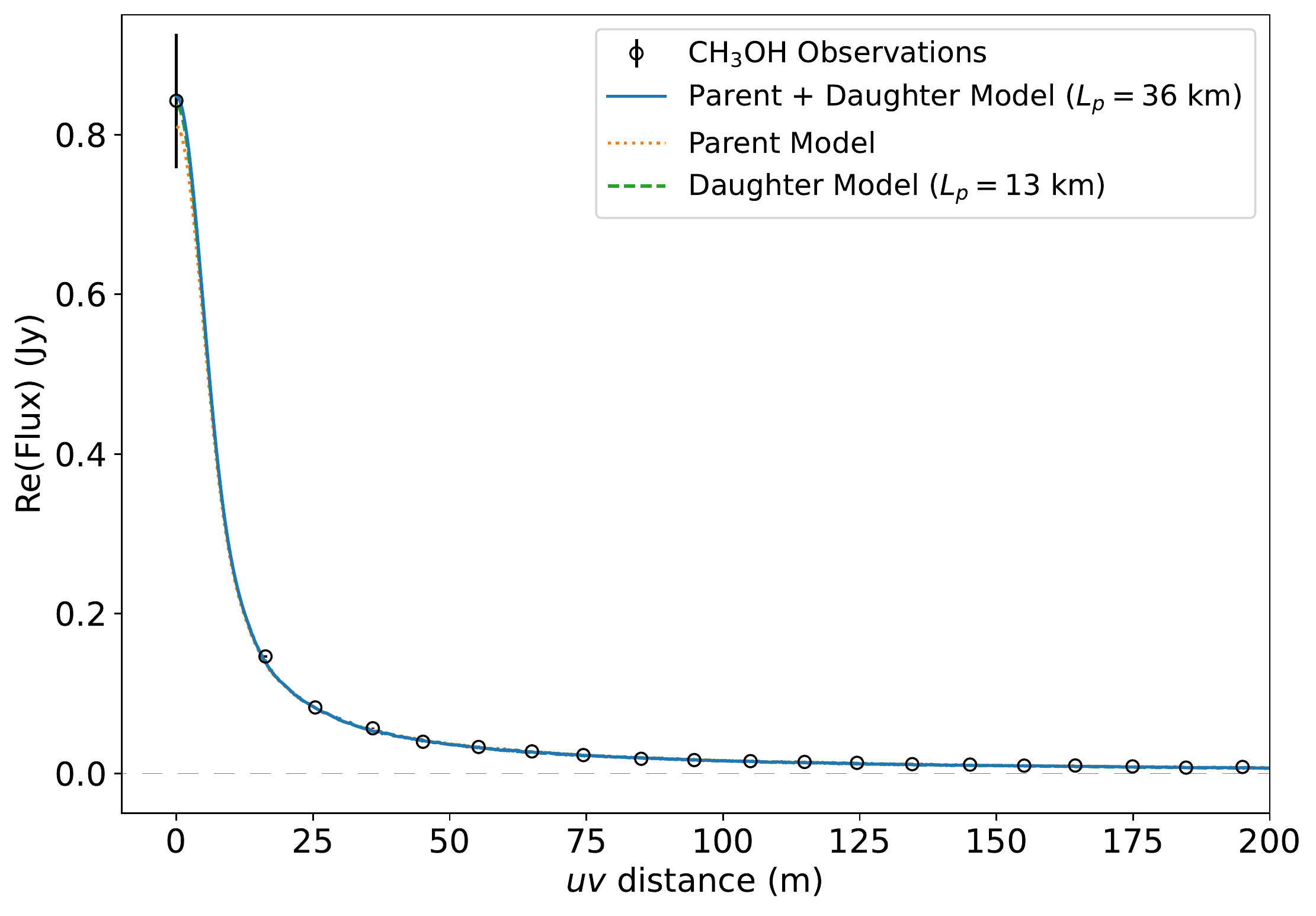}
\includegraphics[height=4cm]{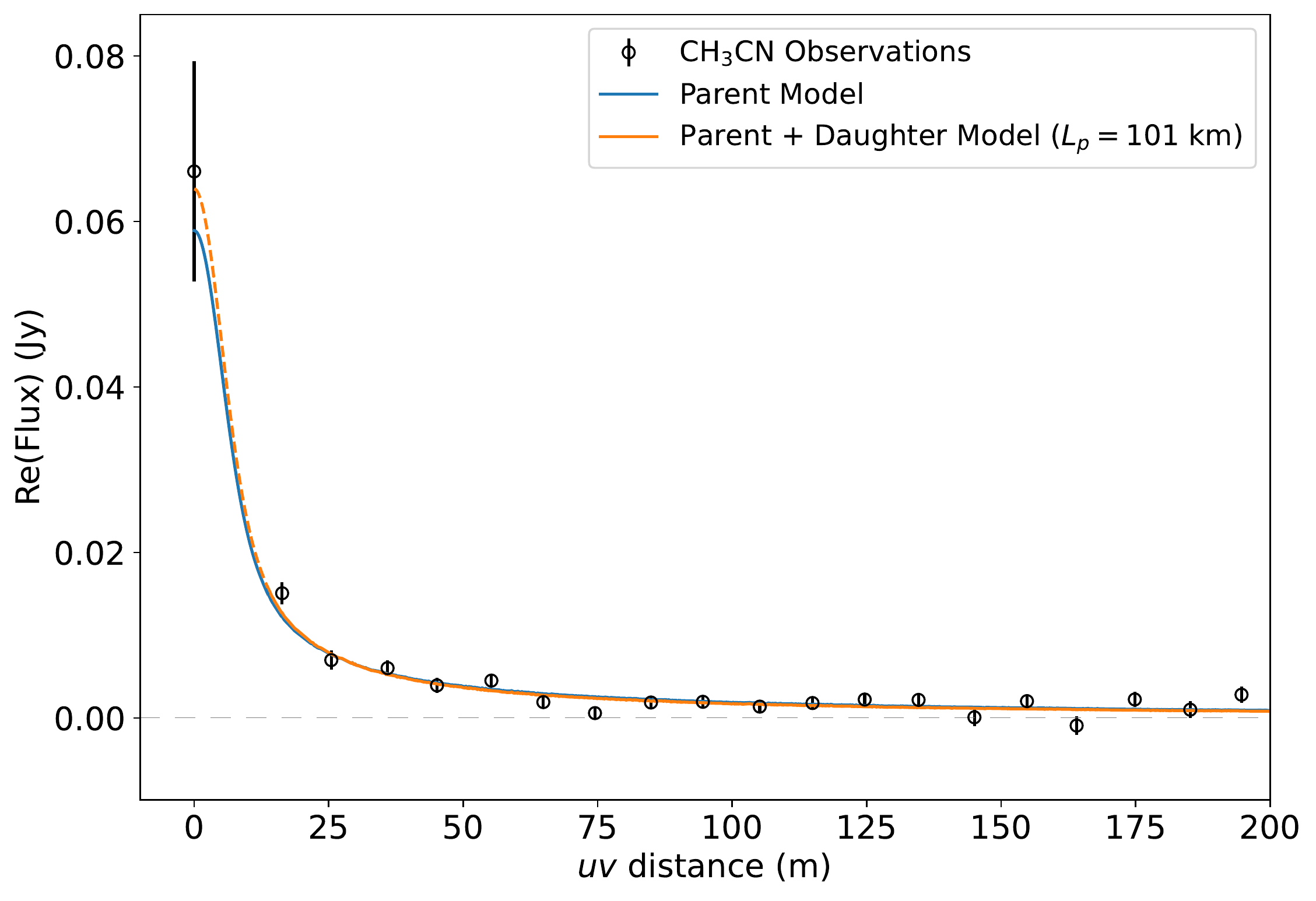}
\caption{Observed interferometric visibilities for HCN, CH$_3$OH and CH$_3$CN, with best-fitting models overlaid. The observations and model data have been (spatially) averaged within 10~m $uv$ bins and (spectrally) averaged over all line emission channels for each species, including 13 lines for CH$_3$OH and 5 lines for CH$_3$CN. The zero-spacing data points (at $uv$ = 0) were taken from the ALMA autocorrelations (observed simultaneously with the interferometric data). Preferred, best-fitting visibility models are shown with solid lines, while dashed and dotted curves show models that are ruled out based on a poorer quality of fit (or other criteria; see text).  \label{fig:parentVis}}
\end{figure}

\begin{figure}
\centering
\includegraphics[height=4.1cm]{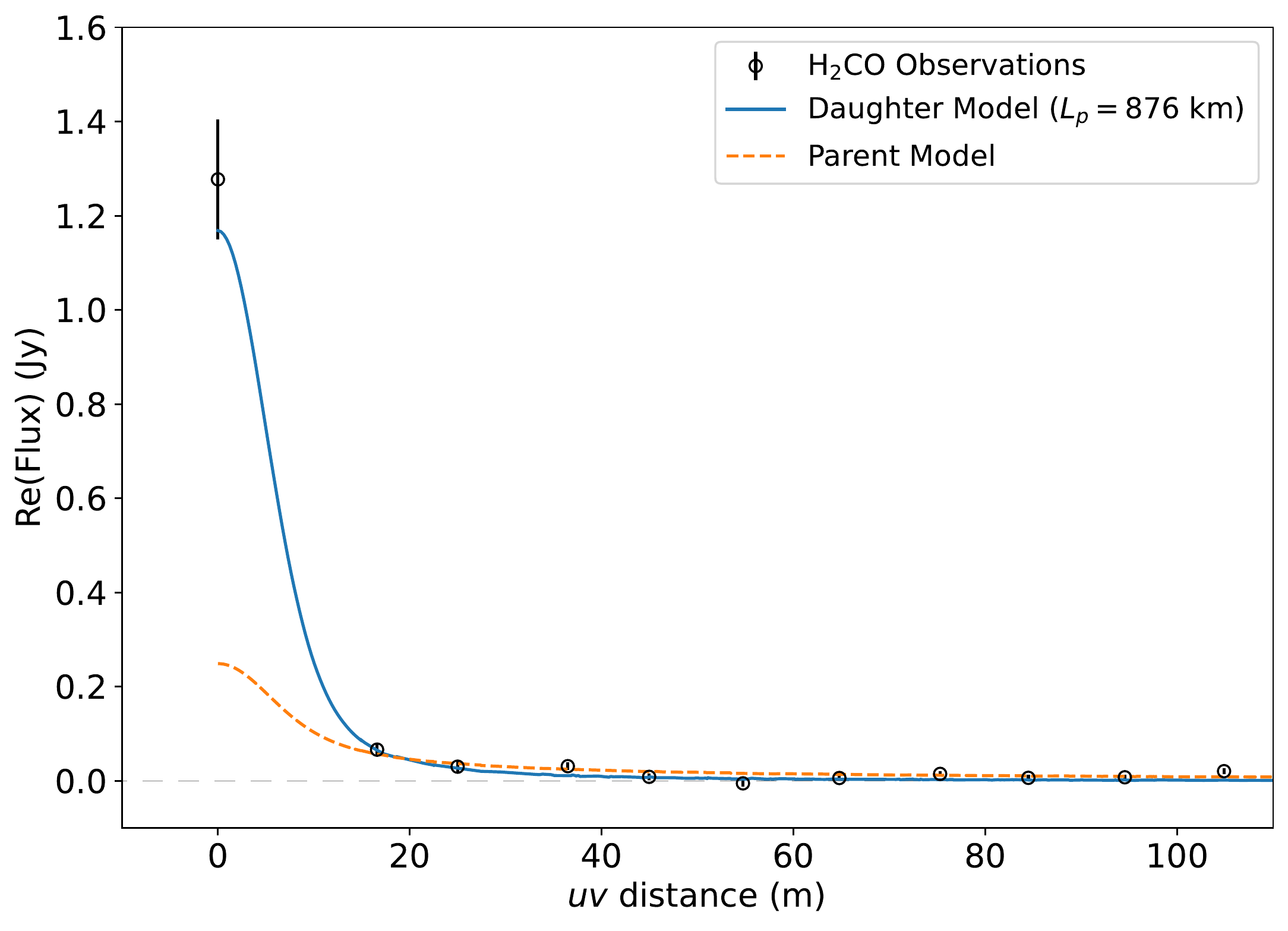}
\includegraphics[height=4.1cm]{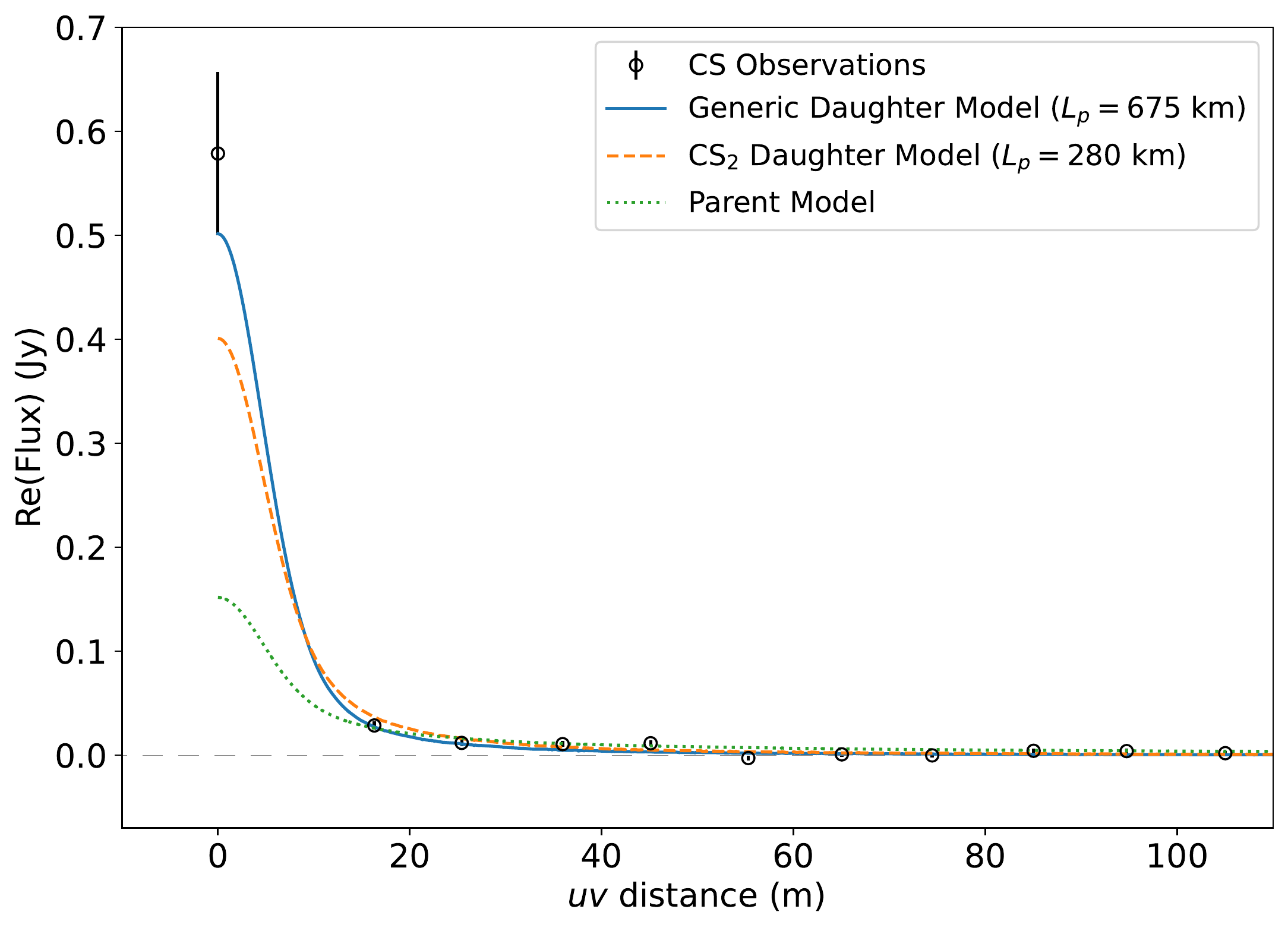}
\includegraphics[height=4.1cm]{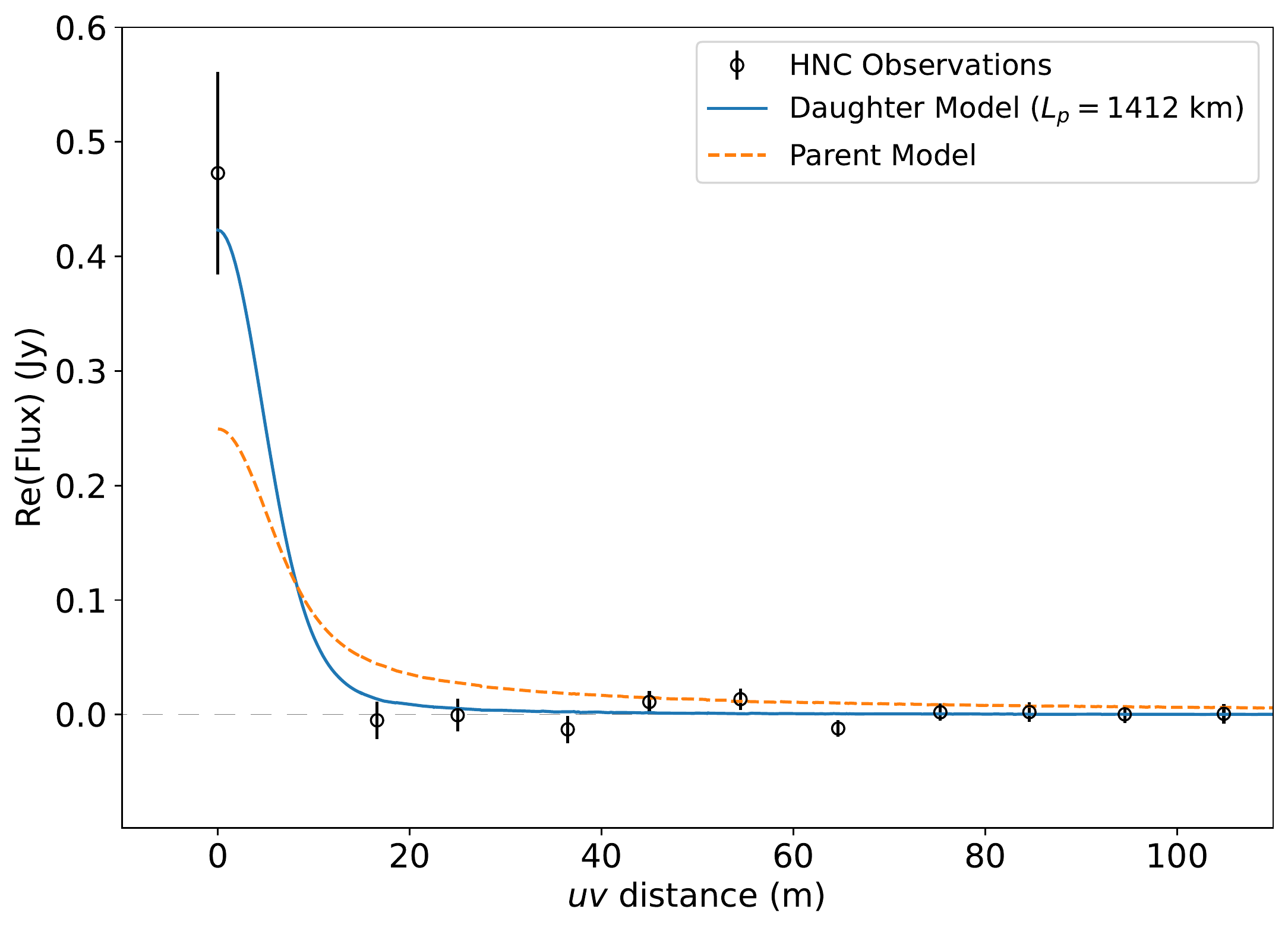}
\caption{Same as Figure \ref{fig:daughterVis}, for H$_2$CO, CS and HNC. \label{fig:daughterVis}}
\end{figure}

To visualize the radial flux distribution for each molecule, in Figures \ref{fig:parentVis} and \ref{fig:daughterVis} we plot the real part of the interferometric visibility as a function of baseline length ($uv$ distance). The observed and modeled fluxes are averaged across the spectral channels containing line emission, and the interferometric data have been further averaged into 10 m-wide $uv$ bins. Statistical error bars on the interferometric data are typically very small due to the large number of data points being averaged together, whereas the total power data points have an additional 10 \% error added in quadrature with the statistical error, to allow for uncertainty in the aperture efficiency of the ALMA antennas. For HCN, a parent outgassing model provides the best fit to the data. In an attempt to better fit the total power ($uv=0$) data point, a daughter distribution of HCN was added, and the model parameters were re-optimized (with variable parent and daughter abundances, and variable HCN parent scale length $L_p$). However, this did not significantly improve the $\chi^2$ value, and the optimized $L_p$ value was found to be $5\pm16$~km, which is consistent with zero, showing that HCN is likely a parent species.

The binned CH$_3$OH visibilities and corresponding model fits are shown in Figure \ref{fig:parentVis} (middle panel). As explained in Section \ref{sec:ch3oh}, in contrast to HCN, a parent + daughter model provides the best fit for CH$_3$OH (with $L_p=36\pm7$~km for the daughter component). The difference between the CH$_3$OH ``parent'' and ``parent + daughter'' models is difficult to see in Figure \ref{fig:parentVis}; unbinned, zoomed CH$_3$OH visibilities are therefore shown in Appendix \ref{sec:visfits}, Figure \ref{fig:ch3ohviszoom}, where the improved fit for the parent + daughter model is evident.  For CH$_3$CN, a parent model fits the data very well. Addition of a CH$_3$CN daughter component slightly improved the fit (resulting in $L_p=101^{+194}_{-101}$~km for the daughter), but the associated drop in $\chi^2$ value corresponded to only $1.6\sigma$, so the improvement in fit quality was not statistically significant.

The binned visibility data for H$_2$CO, CS and HNC are plotted in Figure \ref{fig:daughterVis}, and as shown by the overlaid model curves, these three species can only be well fit using daughter models (their respective best-fitting parent scale lengths are given in Table \ref{tab:results}). To obtain the best visibility fits for these daughter species, we allowed their outflow velocities to differ from those of the underlying H$_2$O distribution (determined from our HCN and CH$_3$OH models), which may be physically justified if they originate from a non-nucleus source. We also attempted to fit these data using `parent only' models, by setting $L_p=0$ and optimizing the abundance to obtain the best fit to the observations. However, in all three cases, a parent model provides a clearly inadequate fit, particularly at the largest scales probed by the total power ($uv=0$) data.  For CS, we also tried to fit the observations assuming production of this species from CS$_2$ photolysis (at a photodissociation rate $\Gamma=2.61\times10^{-3}$, which corresponds to $L_p=280$~km in the jet and $L_p=152$~km in the ambient coma; \citealt{hue15}). As shown by the dashed orange curve, the CS$_2$ photolysis model produces insufficient CS flux on large scales (small $uv$ values) to reproduce the observations. The final, best fitting model parameters for all species are given in Table \ref{tab:results}.

\begin{figure}
\centering
\includegraphics[height=4cm]{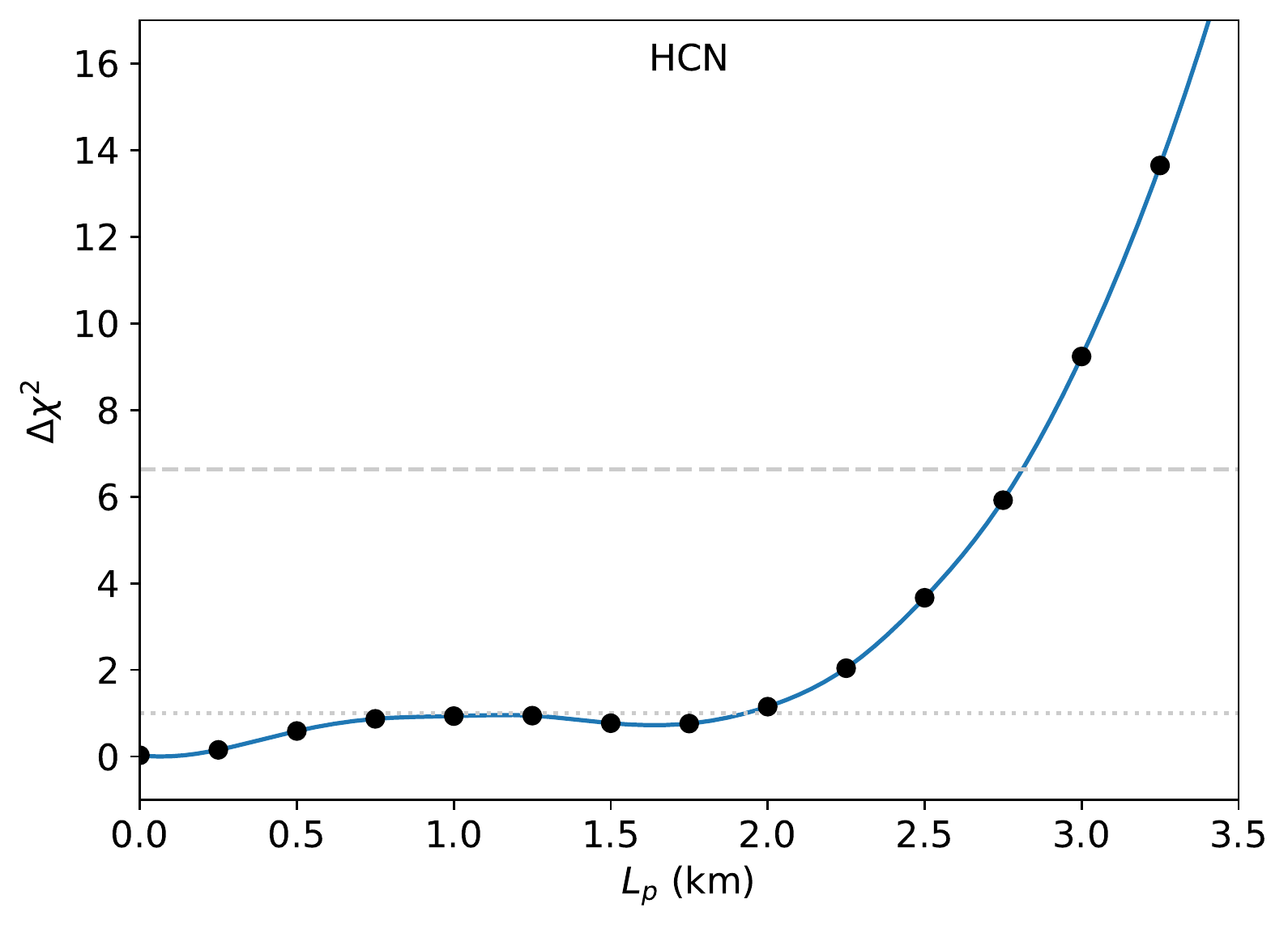}
\includegraphics[height=4cm]{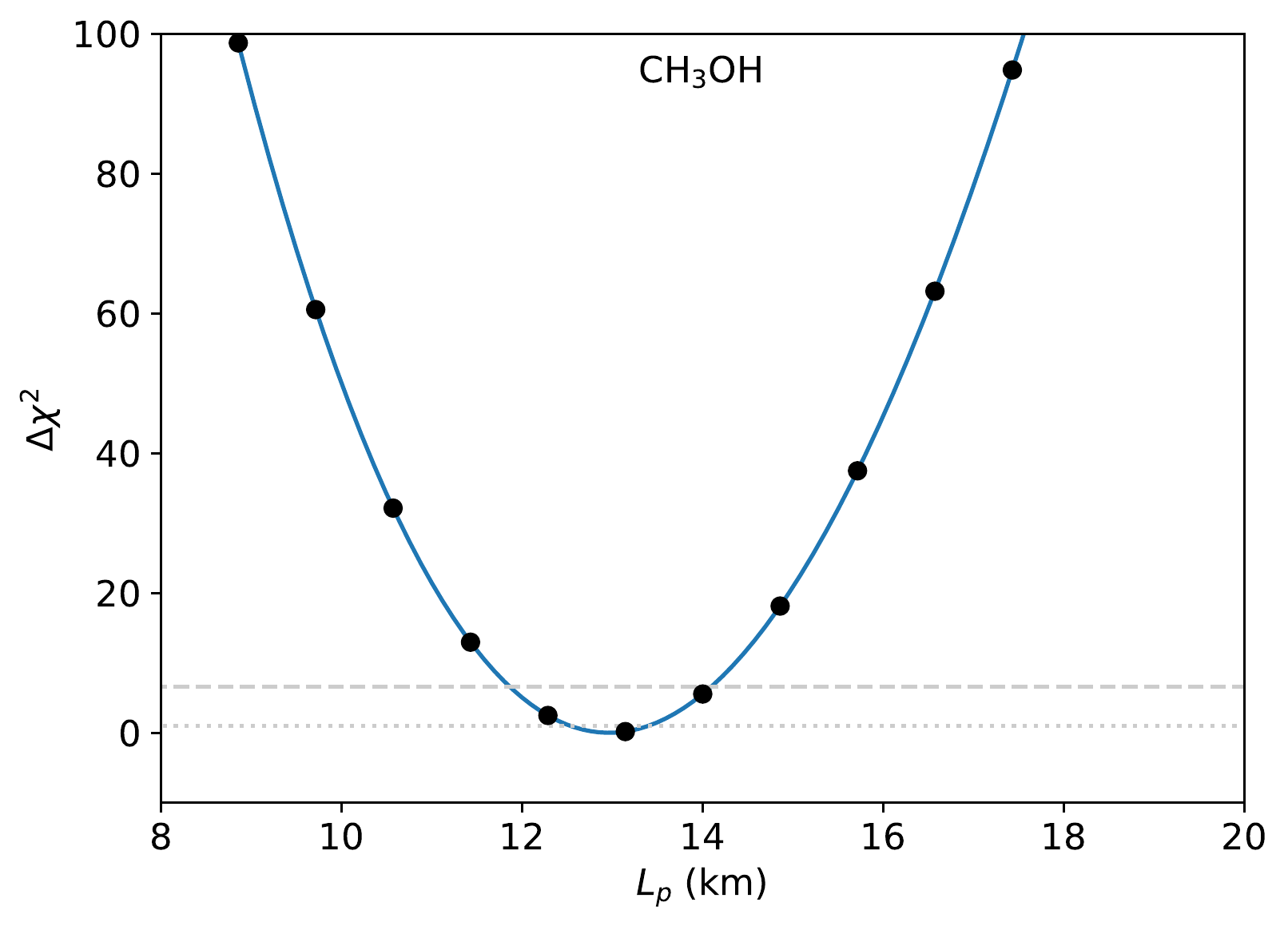}
\includegraphics[height=3.95cm]{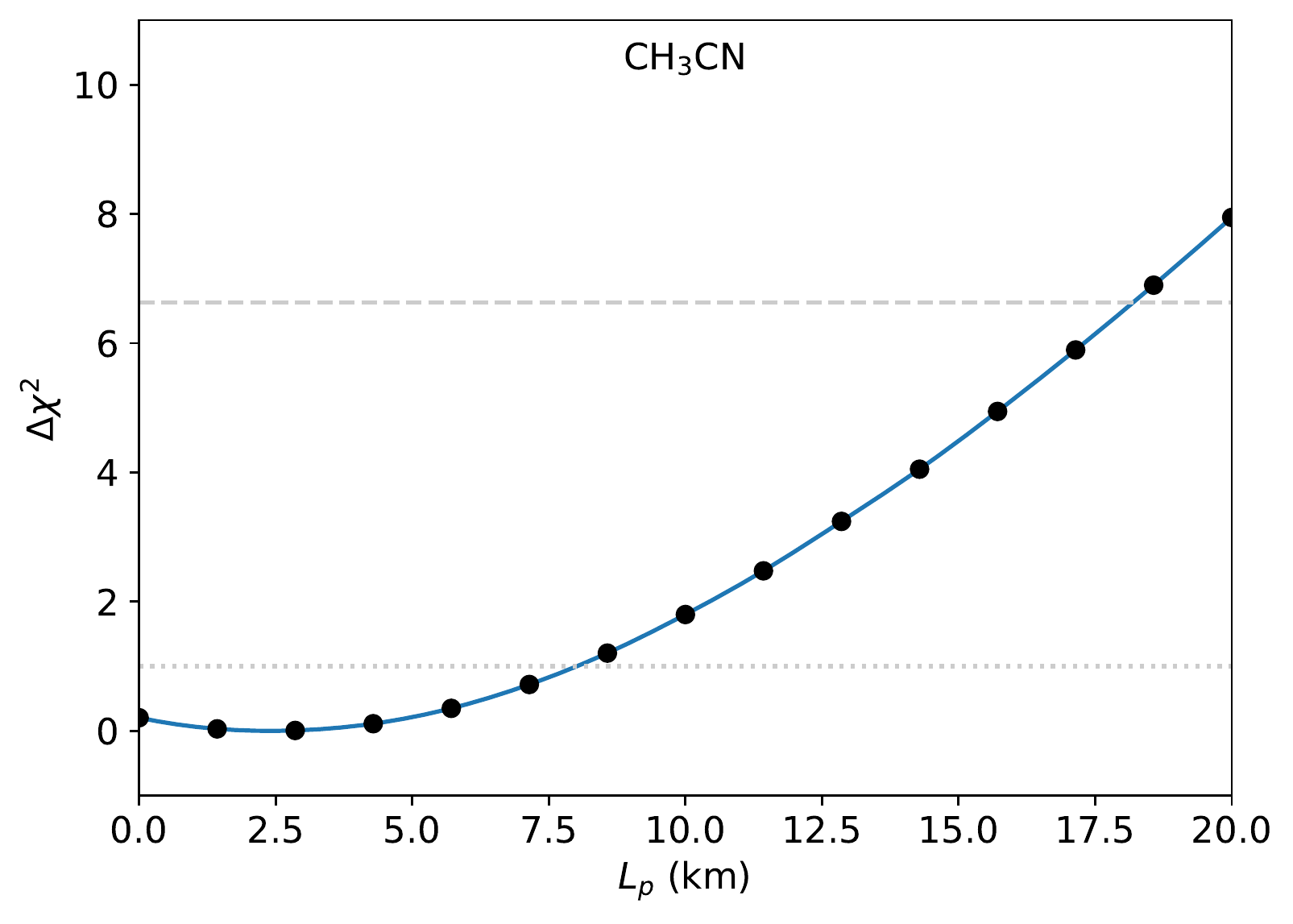}
\includegraphics[height=4cm]{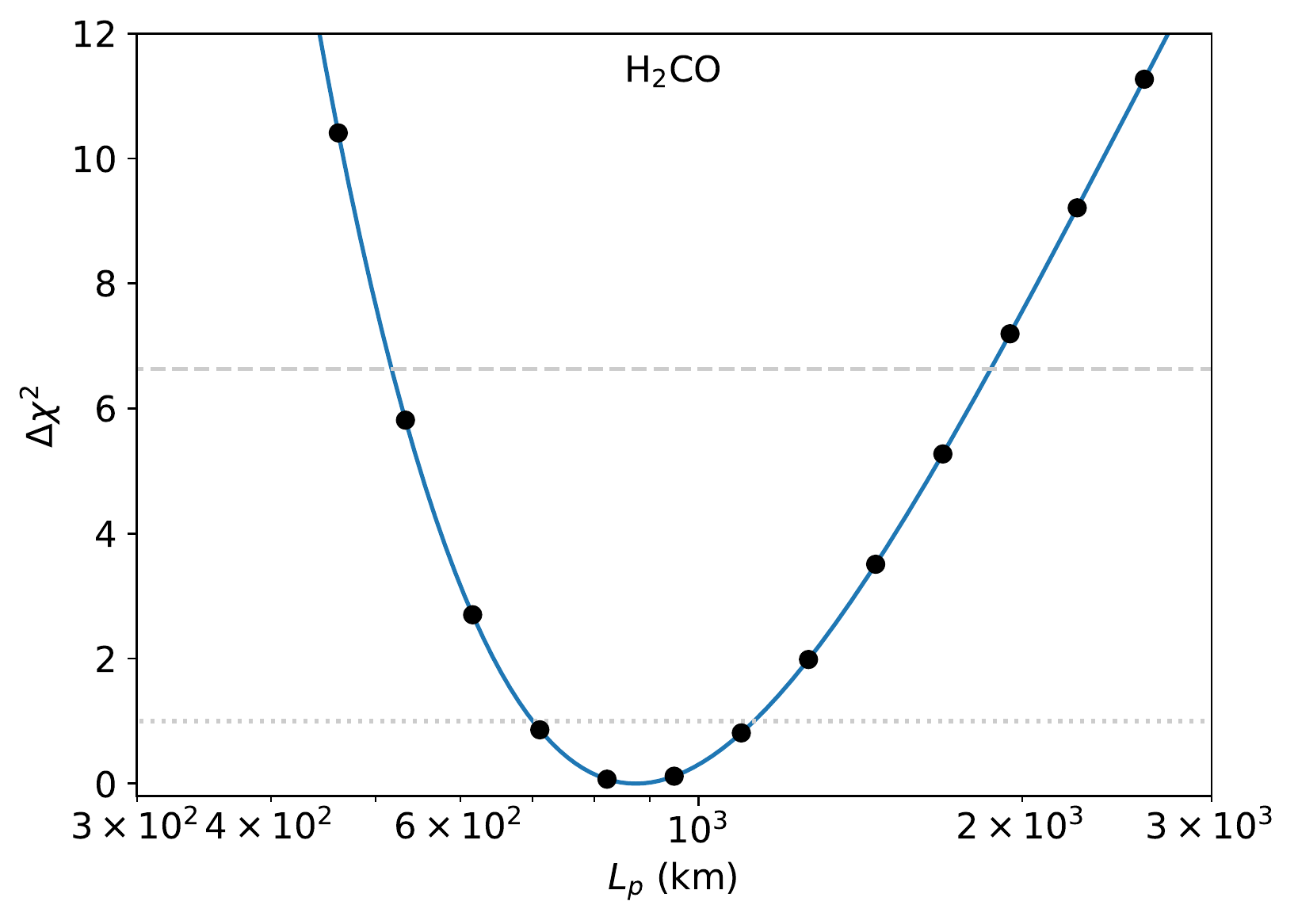}
\includegraphics[height=4cm]{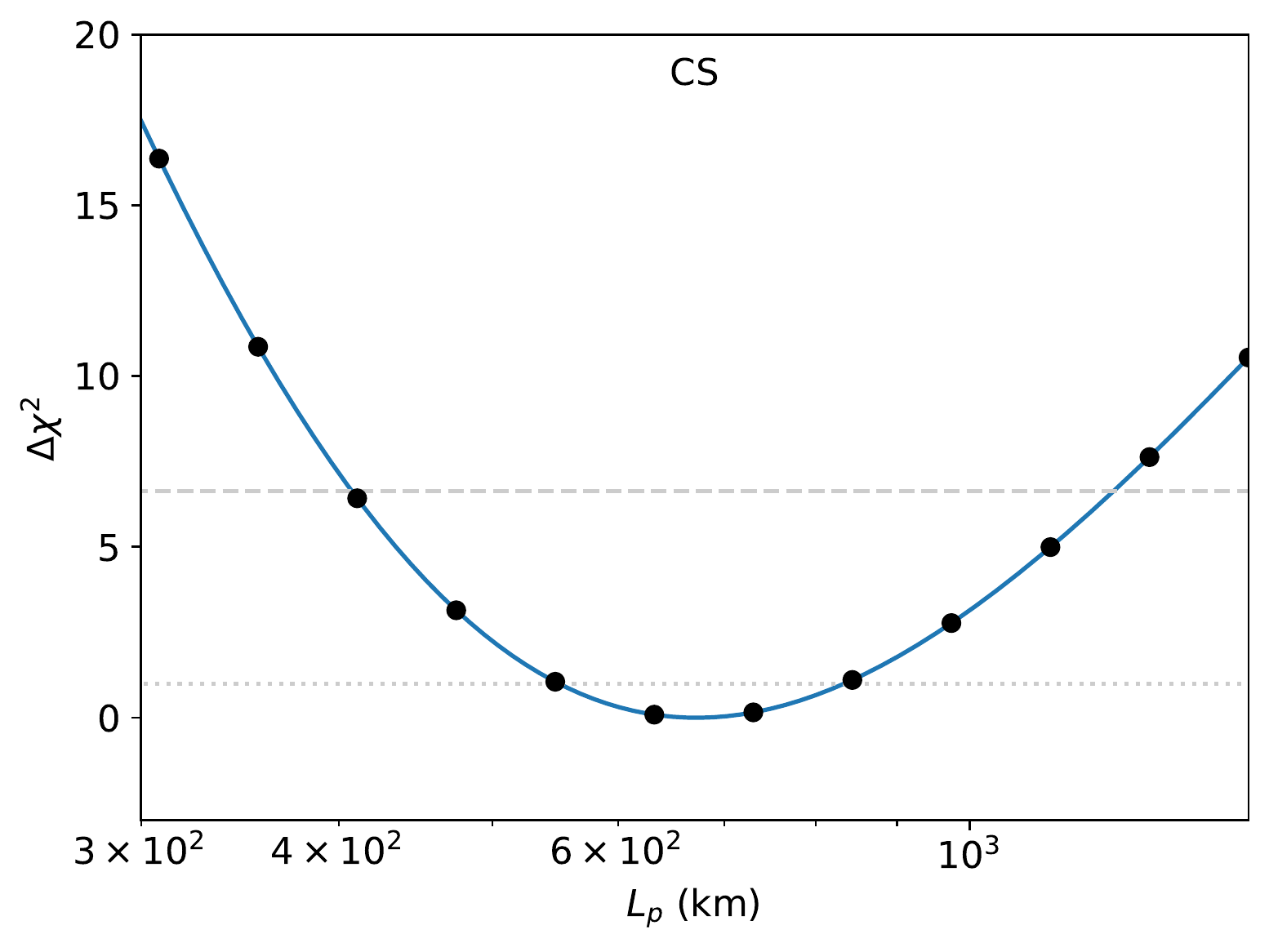}
\includegraphics[height=3.95cm]{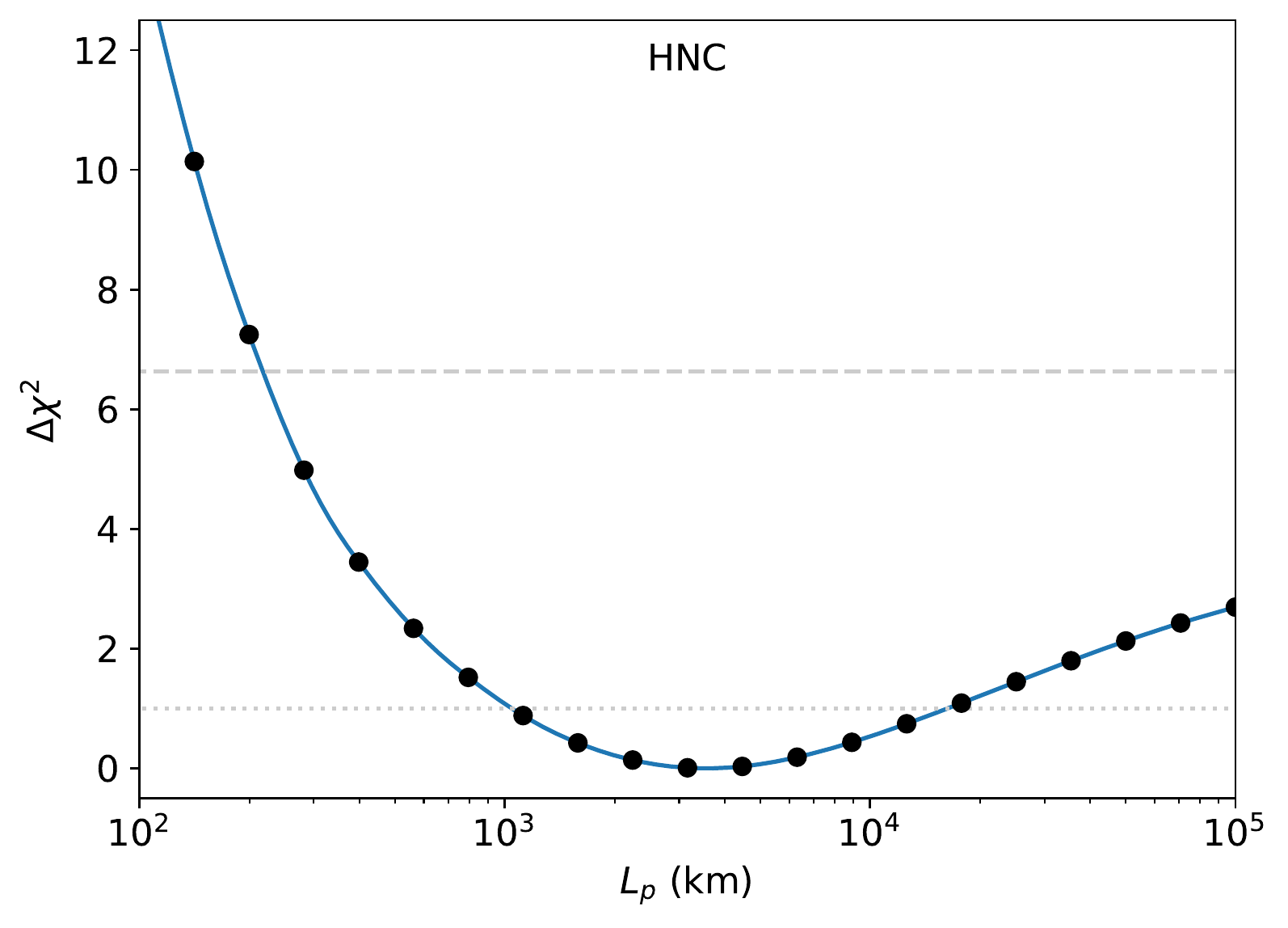}
\caption{Parent scale-length $\chi^2$ analysis for our observed species in 46P, showing the change in chi-square value ($\Delta\chi^2$), as a function of parent scale-length ($L_p$). Dotted horizontal lines show the $\Delta\chi^2=1$ threshold ($1\sigma$), corresponding to 68 \% confidence, and dashed lines show the $\Delta\chi^2=6.63$ threshold ($2.6\sigma$), corresponding to 99 \% confidence. \label{fig:LpChisqs}}
\end{figure}

\begin{table}
\caption{3D visibility modeling results\label{tab:results}}
\begin{center}
\begin{tabular}{llccc}
\hline
\hline
Species&Abundance$^a$ (\%)&$v_1$ (km\,s$^{-1}$)&$v_2$ (km\,s$^{-1}$)&$L_p$ (km)\\
\hline
     HCN & $0.1176^{+0.0003}_{-0.0003}$ & $0.741\pm0.001$ & $0.443\pm0.001$ & $<3$ \\
CH$_3$OH$^b$ & $2.7^{+0.1}_{-0.1}$ & $0.729\pm0.002$ & $0.395\pm0.002$ & 0 / $36\pm7$ \\
CH$_3$CN & $0.011^{+0.002}_{-0.002}$ & $0.729^c$ & $0.395^c$ & $<18$ \\[2mm]
 H$_2$CO & $0.153^{+0.031}_{-0.023}$ & $0.65\pm0.02$ & $0.47\pm0.02$ & $876^{+250}_{-175}$ \\
      CS & $0.022^{+0.004}_{-0.002}$ & $0.66\pm0.02$ & $0.37\pm0.02$ & $675^{+159}_{-124}$ \\
     HNC & $0.0054^{+0.0232}_{-0.0018}$ & $0.69\pm0.05$ & $0.34\pm0.05$ & $1412^{+14,876}_{-367}$ \\
\hline
\end{tabular}
\parbox{0.6\textwidth}{\footnotesize
\vspace*{1mm}
$^a$ Total production rate relative to H$_2$O.
$^b$ CH$_3$OH best-fitting model has both nucleus (parent) and coma (daughter) production; the reported abundance is a sum over both components.\\
$^c$ Held fixed at the CH$_3$OH values (simultaneously observed) due to the low spectral resolution for CH$_3$CN.\\
}
\end{center}
\end{table}

Due to the inverse exponential behavior of cometary molecular densities as a function of radius, uncertainties on the parent scale lengths can be highly asymmetric, and are therefore not adequately represented using the symmetric error bars obtained from the {\tt lmfit} covariance matrix. To address this issue, we calculated the $\chi^2$ surface for each species as a function of $L_p$ (with the abundance as a variable, but other model parameters held fixed), and plotted the resulting $\Delta\chi^2(L_p)$ curve, with cubic spline interpolation between points (with $\Delta\chi^2(L_p) = \chi^2(L_p) - \chi^2_m$, where $\chi^2_m$ is the minimum value). The results are shown in Figure \ref{fig:LpChisqs}, with dotted horizontal lines to show the $1\sigma$ (68 \% confidence) $\Delta\chi^2$ thresholds, and dashed lines to show the $2.6\sigma$ (99 \% confidence) thresholds.

For HCN, the  $\chi^2$ minimum is for $L_p=0$, confirming that this species is a parent. For CH$_3$CN, the smallest $\chi^2$ value occurs around $L_p=2.5$~km, but this minimum is much shallower than $1\sigma$, and is therefore not significant. The CH$_3$CN $\Delta\chi^2$ curve crosses the 99\% confidence threshold at $L_p=18$~km, which represents a strict upper limit on the scale of possible coma production for this species assuming a single single (distributed) source. CH$_3$OH, on the other hand, shows a well defined $\chi^2$ minimum at $L_p=13$~km. However, as explained in Section \ref{sec:ch3oh}, a composite (parent + daughter) model provides the best fit for this species (with a lower $\chi^2$ value than the pure daughter model plotted here); the $L_p$ value of the daughter component in the CH$_3$OH composite model is well constrained as $36\pm7$~km, so this is reported in Table \ref{tab:results} instead.

Assuming a scale length of $L_p=36$~km for a sublimating icy grain source (\emph{i.e.} the same as derived for CH$_3$OH), the abundance upper limit (at the 99\% confidence level) for icy grain production of HCN is $<0.001$ \% relative to H$_2$O. This corresponds to $<1$\ \% of the total HCN. The associated upper limit for icy grain production of CH$_3$CN is $<0.01$ \%, which is comparable to the measured CH$_3$CN parent abundance, and shows that a significant source of CH$_3$CN production from icy grains in the coma, close to the nucleus, cannot be ruled out by our data.

H$_2$CO, CS and HNC all exhibit well-defined $\chi^2$ minima at substantially larger $L_p$ values than the other observed species, so these molecules are confirmed as originating primarily in the coma of comet 46P, rather than from the nucleus. While the H$_2$CO and CS $L_p$ values are well constrained (at $876^{+250}_{-175}$~km and $675^{+159}_{-124}$~km, respectively), the HNC uncertainty interval is much larger and highly asymmetric, with an $L_p$ value lying between 1045~km and 14,880~km (at $1\sigma$ confidence). At the stricter, 99 \% confidence level, the $L_p$(HNC) value is only constrained by a lower limit of $>218$~km.

%Note: H2CO OPR assumed to be 3. CH3CN A/E ratio assumed to be 1. CH3OH A/E ratio found to be consistent with 1.

\section{Discussion}

\subsection{HCN}

The high spatial resolution of ALMA combined with the close geocentric distance of 46P places unusually tight constraints on the production scale length of HCN. With an upper limit of $L_p<3$~km, it is likely that HCN originates primarily from outgassing by the nucleus of the comet. At the outflow velocity of 0.74~km\,s$^{-1}$ (in the sunward jet), any HCN daughter production would need to occur on a timescale of $<4$~s, which is significantly shorter than the lifetimes of any known coma molecules (typically in the range $10^3$--$10^6$~s; \citealt{hue15}). Our scale length upper limit is consistent with (although significantly better constrained than) the previous upper limits of $L_p<50$~km in C/2012 F6 (Lemmon) and C/2012 S1 (ISON) \citep{cor14} and $L_p<100$~km in C/2015 ER61 \citep{rot21}, which were obtained using ALMA and Atacama Compact Array (ACA) data. Including the IRAM Plateau de Bure interferometric study of comet Hale-Bopp \citep{win97} and the Keck NIRSPEC study of C/2014 Q2 (Lovejoy), the body of evidence demonstrating HCN release from direct sublimation of nucleus ices, with no clear evidence for production in the coma, is now substantial. Coma HCN abundance measurements in fully-activated comets thus provide a valid proxy for the abundance of this molecule in the nucleus ices.

Our derived HCN abundance is $0.1176\pm0.0003$ \% with respect to H$_2$O, with the error bar accounting only for statistical uncertainties. Adopting a conservative estimate for the ALMA flux scale uncertainty of 10 \%, and an extra uncertainty of 10 \% in the H$_2$O production rate (added in quadrature), the abundance error increases to $\pm0.016$ \%. Our measured HCN abundance in comet 46P is therefore consistent with the value of $0.11\pm0.01$ \% obtained by \citet{biv21} within the same month, and matches the average value observed previously in OCCs and JFCs using radio spectroscopy \citep{boc17}.

\subsection{CH$_3$OH}
\label{sec:dis_ch3oh}

Our observations reveal the presence of two coexisting outgassing modes for CH$_3$OH, consisting of (1) production directly from the nucleus and (2) production from a near-nucleus coma source. Our best-fitting SUBLIME model indicates that both sources are responsible for similar amounts of CH$_3$OH production, with best-fitting abundances (with respect to H$_2$O) of $1.19\pm0.13$ \% for the nucleus (parent) source of CH$_3$OH gas and $1.18\pm0.13$ \% for the distributed (coma) source.

The presence of icy grain sublimation has been known to be an important source of gases in cometary comae since the EPOXI mission to comet 103P/Hartley 2 \citep{ahe11}. Using infrared spectroscopy, \citet{pro14} observed 1~$\mu$m sized water ice grains in the coma of 103P, while \citet{kel13} observed a halo of larger, longer-lived ice chunks (with sizes $\geq1$~cm) surrounding the nucleus at distances up to $\sim40$~km. Using a detailed DSMC coma model, \citet{fou13} determined that 77 \% of the H$_2$O outgassing in 103P originates from sublimation of icy grains in the coma. Such an excess of H$_2$O production compared with its small nucleus size, led to 103P being designated as a `hyperactive' comet. 

46P/Wirtanen is also hyperactive \citep{lis19,mou23}, and the large H$_2$O production rate for its nucleus size can, at least in part, be attributed to the presence of icy grains in the coma \citep{bon21}. \citet{kni21} explained the different OH and CN morphologies in 46P as being partly due to OH production from an icy-grain H$_2$O source. However, \citet{pro21} exclude pure H$_2$O ice grains in the coma of this comet. Based on the non-detection of H$_2$O ice absorption features, these authors suggested two possible explanations for the hyperactivity of 46P: icy grains ($\sim1$ $\mu$m in size) containing a small amount of low-albedo dust, or larger chunks ejected from the nucleus, containing significant amounts of water ice.  Previous ALMA observations by \citet{cor17b} identified nucleus outgassing as the primary source of CH$_3$OH in the coma of C/2012 K1 (PanSTARRS). On the other hand, a distributed CH$_3$OH source (with a scale length $\sim550$~km) was observed using the James Clerk Maxwell Telescope (JCMT) in comet 252P/LINEAR, and was attributed to CH$_3$OH production from icy grains in the coma \citep{cou17}. It is therefore reasonable to attribute the source of our observed CH$_3$OH daughter distribution in 46P (around 40 km from the nucleus) to the presence of sublimating icy grains in the coma, composed of a mixture of CH$_3$OH, H$_2$O and dust. The sublimation rate of coma icy grains depends on their size and composition. Based on the calculations of \citet{bee06}, and assuming that the gas and dust outflow velocities are coupled, our derived lifetime of $496\pm97$~s for the CH$_3$OH-producing grains in 46P implies a dirty ice grain size of $\sim10$~$\mu$m. If the dust is moving outward slower than the gas, then the grains could be larger.

The presence of sublimating icy grains can significantly impact the coma temperature, as shown by the direct simulation Monte Carlo (DSMC) models of \citet{fou12} and \citet{fou14}, due to the transfer of excess kinetic energy from sublimated molecules to the surrounding gas. In the absence of such a mechanism, the gas kinetic temperature in coma physical models \citep[\emph{e.g.}][]{rod04,ten08} falls rapidly with distance from the nucleus, from $\gtrsim100$~K at $r=0$ to $\sim10$~K at $r=100$~km, due to quasi-adiabatic expansion. Such steeply declining temperatures are at odds with coma observations. For example, the observations of comets 73P-B/Schwassmann–Wachmann, 103P/Hartley 2, C/2012 S1 and 46P/Wirtanen \citep{bon08,bon13,bon14,bon21} and C/2012 K1 \citep{cor17b} revealed a shallower temperature decay --- and in some cases, increasing temperatures --- as a function of radius, attributable to sublimative heating. Spatially resolved temperature observations therefore provide an indirect probe for the presence of icy grains in cometary comae.

Our derived 46P coma kinetic temperature profile (Figure \ref{fig:Tprof}) shows an initial increase with radius on the sunward side of the nucleus, followed by a relatively slow decrease, consistent with strong coma heating due to the sublimation of icy grains within a few hundred kilometers of the nucleus. On the antisunward side, the temperature drop is also less steep than predicted by hydrodynamical and DSMC models for other comets, and flattens out to a higher temperature than predicted without the presence of icy grain heating \citep{rod04,ten08,fou12}. This temperature behavior therefore suggests the presence of significant icy grain sublimation in the 46P coma, consistent with conclusions based on spatially resolved infrared observations of H$_2$O in this comet \citep{bon21}. Dedicated theoretical modeling will be required to further investigate this hypothesis.

Our total CH$_3$OH abundance of $2.7\pm0.1$ \% including the nucleus and coma sources is consistent (at the $2\sigma$ level) with the value of $3.03\pm0.23$ \% observed by \citet{bon21} on 2018-12-18, but somewhat smaller than the value of $3.38\pm0.03$ \% observed by \citet{biv21} during December 11--18, and larger than the value of $1.6\pm0.1$ \% observed by \citet{ber22} on Dec. 22-28. These differences could be explained by temporal variability of the CH$_3$OH and H$_2$O outgassing rates, since significant short-timescale variations in $Q({\rm CH_3OH})$ (over a period of several hours to several days) were observed in this comet by \citet{rot21b}. Indeed, our CH$_3$OH production rate of $(1.9\pm0.1)\times10^{26}$\, s$^{-1}$ on 2018-12-07 is very close to the value of $(2.2\pm0.1)\times10^{26}$\, s$^{-1}$ observed on 2018-12-06 by \citet{kha23}, and $(2.1\pm0.1)\times10^{26}$\, s$^{-1}$ on 2018-12-11 \citep{biv21}, demonstrating good consistency between the ALMA, IRTF and IRAM measurements in this case.

The presence of an icy grain source for CH$_3$OH, but not HCN implies that the icy grains are chemically distinct from the bulk (sublimating) ice within the nucleus. \citet{dra12} also inferred the presence of icy grain sources of CH$_3$OH and HCN in comet 103P, and concluded that CH$_3$OH is more intimately mixed with H$_2$O ice than HCN, with a larger abundance of CH$_3$OH in the icy grains. The apparent dichotomy between the abundances of CH$_3$OH and HCN in the two cometary ice storage reservoirs (icy grains \emph{vs.} bulk nucleus) was interpreted by \citet{dra12} as due to thermal evolution of the nucleus. A primordial origin in the protoplanetary disk or interstellar cloud, prior to accretion of the comet, is also possible. Spatial differentiation of O and CN-rich molecules between gas and ice phases has been observed in disks \citep{ber18,boo21,obe21}, indicating an active carbon chemistry occurs in the gas phase, while O-bearing species such as CH$_3$OH and H$_2$O remain largely frozen on grains. Non-uniform mixing of these distinct chemical reservoirs during comet accretion could lead to the observed spatial heterogeneity of the CH$_3$OH/HCN ratio within the nucleus.

\subsection{CH$_3$CN}

A production scale length upper limit of $L_p<18$~km was derived for CH$_3$CN, based on our visibility modeling. The ALMA data show no significant evidence for production of this molecule as a coma daughter species (either from photolysis of a parent molecule, or from sublimation of icy grains), so we conclude that the primary source of CH$_3$CN is from direct sublimation of molecular ices in the nucleus. The abundance of this molecule ($0.011\pm0.002$ \%) is consistent with, but toward the lower end of the range of values (0.008--0.054 \%) previously observed in comets \citep{boc17}.

This is the most complex nitrile confirmed to be present in cometary ice to-date. It has long been detected in the gas phase in the interstellar medium and protostellar envelopes \citep{her09}, while recent studies using ALMA have found CH$_3$CN to be widespread in protoplanetary disks where it is believed to be formed primarily \emph{via} chemistry on grain surfaces \citep{obe15,ber18,ile21}, before being thermally-desorbed into the gas phase where it can be observed. CH$_3$CN has not been detected so-far in interstellar ices, but our measured abundance in 46P is consistent with the upper limit of $\lesssim2$ \% (with respect to H$_2$O) found recently along two interstellar sightlines using the James Webb Space Telescope \citep{mcc23}.

The CH$_3$CN/HCN ratios we measure in 46P are consistent with, or somewhat higher than the values  observed in nearby protoplanetary disks by \citet{ber18} and \citet{ile21}, implying a likely genetic relationship between protoplanetary disk and cometary nitriles. Our observed CH$_3$CN/H$_2$O ratio also matches that found in the (gas plus ice) phase of the GM Aur planet-forming disk by \citet{ile21}, but is $\sim8$ times lower than that found in the AS 209 disk. The utility of such comparisons is limited, however, due to the action of gas-phase chemical processes, which can modify the CH$_3$CN/HCN/H$_2$O ratios found in the disk gas compared with those in the ice. Cometary observations therefore remain as a perhaps more useful probe of the abundances of complex nitriles in the ice reservoir of our solar system's planet-forming disk.

\subsection{H$_2$CO}

Our measured H$_2$CO production scale length of $L_p=876^{+250}_{-175}$~km at $r_H=1.07$ is compatible with the values measured previously using ALMA of $2200^{+1100}_{-800}$~km at $r_H=1.17$~au in C/2015 ER61 (PanSTARRS) \citep{rot21b}, $1200^{+1200}_{-400}$ at $r_H=1.47$ au in C/2012 F6 (Lemmon) and $280\pm50$ at $r_H=0.54$ au in C/2012 S1 (ISON) \citep{cor14}. When scaled by $r_H^{-2}$, these $L_p$ values are in reasonably close agreement, consistent with H$_2$CO production by photodissociation of a (molecular) parent species at a rate $\sim7\times10^{-4}$~s$^{-1}$ (at $r_H=1$ au), in an optically thin coma. However,  \citet{biv99} measured a much larger $L_p$ value of $7000r_H^{-1.5}$~km in comet C/1996 B2 (Hyakutake) using a single-dish radio telescope, and \citet{mei93} derived $L_p\sim3600$~km in 1P/Halley at $r_H=0.89$ au \emph{via} in-situ mass spectrometry.  CH$_3$OH photolysis is expected to produce H$_2$CO in the outer coma, but cannot be responsible for the observed H$_2$CO in these comets, since it occurs at a rate of $1.0\times10^{-5}$~s$^{-1}$ at $r_H=1.06$~au (\citealt{hue15}; see also \citealt{hea17}), which is almost two orders of magnitude smaller than the H$_2$CO parent photodissociation rate required to fit our ALMA observations. Indeed, the corresponding CH$_3$OH dissociation scale in the 46P sunward jet is 73,000~km, which is much larger than the maximum angular scale of 1400~km spanned by the ALMA primary beam FWHM. Alternative H$_2$CO parents must therefore be sought.

The absence of other known C,H,O-bearing molecules with sufficient abundances has led to the idea that H$_2$CO may be released in the coma from the thermal breakdown of organic-rich dust particles. Formaldehyde polymer (or polyoxymethylene; POM) was found to provide a plausible explanation for the observed H$_2$CO parent scale length in comet 1P/Halley \citep{cot04}, and for the heliocentric dependence of H$_2$CO production rates observed in O1/Hale-Bopp \citep{fra06}. The dissociation scale length of solid-phase POM depends strongly on the particle size and temperature \citep{fra06}, so the observed variations in $L_p({\rm H_2CO})$ can be explained as a result of differing size distributions for the POM-rich dust grains in the different comets observed to-date.  Conversely, \citet{mil06} noted a lack of evidence for POM in Giotto mass spectra of comet Halley, and argued that it constitutes an unlikely source of H$_2$CO in the coma. Furthermore, mass spectrometry by the Rosetta mission found no evidence for POM in comet 67P \citep{han22}, so alternative explanations may be required to explain the distributed H$_2$CO source. Moreover, given the limited sample size and large disparity between the observed $L_p({\rm H_2CO})$ values, it will be important to conduct more observations of H$_2$CO distributions in different comets, over a range of coma size scales, at differing heliocentric distances, to better characterize the behavior of the H$_2$CO source(s) and help constrain the properties of its still-elusive parent.

By simultaneously fitting the H$_2$CO parent scale length and production rate ratio relative to water, we derive an H$_2$CO abundance of $x=0.153^{+0.031}_{-0.023}$ \%, which is larger than the value of $x=0.38\pm0.02$ \% obtained from IRAM 30-m observations by \citet{biv21} using $L_p({\rm H_2CO})=5000$~km. This discrepancy is not surprising considering the retrieved H$_2$CO production rate from single-dish observations scales with the adopted $L_p$ value. Using $L_p=880$~km, the IRAM data are consistent with $x=0.13\pm0.01$ \%. The H$_2$CO production rate of $(9.1\pm0.9)\times10^{24}$~s$^{-1}$ obtained by \citet{cou20} using the JCMT assuming $L_p=860$~km, also matches our derived H$_2$CO production rate. Our H$_2$CO abundance is the second-lowest value reported in a comet to-date, at radio wavelengths \citep{boc17}, but it should be remembered that the majority of those values were obtained using assumed (rather than directly measured) $L_p({\rm H_2CO})$ values. We therefore emphasize the importance of using an accurate parent scale length when deriving H$_2$CO production rates. 

An upper limit of $x<0.064$ \% was obtained by \citet{bon21} using Keck infrared observations, but this was calculated assuming H$_2$CO release directly from the nucleus (\emph{i.e.} $L_p=0$) --- a common assumption among infrared spectroscopists \citep[\emph{e.g.}][]{dis06,del16b} due to the difficulty of deriving $L_p$ values from infrared spatial profiles --- which inevitably underestimates the total coma H$_2$CO abundance in the presence of distributed sources.  Our ALMA observations show no evidence for H$_2$CO production by the nucleus, with a ($3\sigma$) upper limit of 0.08 \% on the abundance of any parent H$_2$CO, consistent with the \citet{bon21} result.  The relatively short H$_2$CO parent lifetimes derived using ALMA imply significant daughter production of H$_2$CO in the near-nucleus coma that could, in some situations, appear similar to a parent source. Further investigations of H$_2$CO spatial emission profiles at the angular scales of a few arcseconds probed by infrared spectroscopy are therefore warranted,

Despite the evidence for a lack of H$_2$CO in cometary ices inferred using ALMA observations, it remains one of the most widespread gas-phase molecules in the Galaxy. Abundances of H$_2$CO relative to water can reach several per-cent in warm protostellar gas \citep{cec00,ehr00}, while mid-infrared spectroscopy indicates possible abundances $\sim6$ \% in ices around low-mass protostars  \citep{boo15}.  The lack of detectable H$_2$CO in cometary nuclei using ALMA therefore provides evidence for chemical processing to destroy H$_2$CO ice in (or during its passage to) the protosolar disk, which is expected to occur as part of the pathway to forming more complex organic molecules (including biologically relevant species), starting with hydrogenation of H$_2$CO to make CH$_3$OH ice \citep{chu16}. Although formaldehyde is commonly detected in the gas-phase in protoplanetary disks \citep[\emph{e.g.}][]{peg20}, total H$_2$CO/H$_2$O masses in the range $\sim10^{-5}$--$10^{-3}$ were recently measured in a sample of five disks by \citet{guz21}, so relatively low H$_2$CO abundances in cometary nuclei ($\lesssim6\times10^{-4}$, as found by our study and \citealt{bon21}), may not be surprising. 

Of relevance to the nature of the H$_2$CO parent, it is noteworthy that the H$_2$CO outflow velocity of $0.65\pm0.02$~km\,s$^{-1}$ in the sunward jet is significantly less than the HCN (parent molecule) outflow velocity of $0.74$~km\,s$^{-1}$ measured only 2 hours earlier. Such a rapid drop in the jet velocity over this period seems unlikely, so this result may constitute evidence that the H$_2$CO parent is flowing radially outward at a slower rate than the gases sublimating directly from the nucleus. A similar effect is also observed for the daughter species CS (observed simultaneously with CH$_3$OH) as well as HNC (albeit, at lower confidence). The outflow velocities of coma dust grains are significantly lower than that of the gas particles due to the mass disparity between these fluids, which causes the dust to lag behind the gas \citep{cri04}. A relatively low outflow velocity for H$_2$CO, CS and HNC may therefore indicate production of these molecules from a slower-moving dust precursor.

\subsection{CS}

The CS radical in cometary comae is commonly believe to originate from photodissociation of the CS$_2$ parent molecule \citep{jac82,rod06,fel10}. Our derived CS parent scale length of $L_p=675^{+159}_{-124}$~km, at an outflow velocity of 0.73~km\,s$^{-1}$ would therefore correspond to a photodissociation rate $\Gamma({\rm CS_2})=(1.08\pm0.20)\times10^{-3}$~s$^{-1}$ at $r_H=1.06$~au (or $(1.21\pm0.22)\times10^{-3}$~s$^{-1}$ at $r_H=1$~au). This value is consistent (within errors) with the value derived by \citet{fel99} using spatially resolved HST STIS imaging of CS in comet C/1999 H1 (Lee). However, our inferred $\Gamma({\rm CS_2})$ value is significantly smaller than the published rates: $2.9\times10^{-3}$~s$^{-1}$ \citep{hue15}, $1.9\times10^{-3}$~s$^{-1}$ \citep{hea17} and $1.7\times10^{-3}$~s$^{-1}$ \citep{jac86}, obtained using experimentally-derived CS$_2$ photodissociation cross sections. Even if only one of these rates is correct, the smaller value for the CS parent photodissociation rate obtained by our study implies that CS$_2$ photolysis cannot be the main source of CS in the coma of 46P. This is similar to the conclusion of \citet{rot21}, who derived a CS parent photolysis rate (at $r_H=1$~au) of $(3.6^{+5.4}_{-2.2})\times10^{-4}$~s$^{-1}$ in comet C/2015 ER61, which is even smaller than our value derived for 46P, although the rates remain consistent at the $2\sigma$ level. \citet{biv22} also derived a small CS parent photolysis rate of (4--8)$\times10^{-4}$~s$^{-1}$ in C/2020 F3 (NEOWISE).

After considering the possible CS sources in comet C/2015 ER61, \citet{rot21} determined that H$_2$CS was a possible parent for CS, but noted that the abundance of H$_2$CS in many comets is actually less than the CS abundance, so it cannot be the parent of all cometary CS. Indeed, the upper limit on the H$_2$CS abundance in comet 46P of $<0.016$ \% \citep{biv21} is at odds with this hypothesis based on our derived CS (parent) abundance of $0.022^{+0.004}_{-0.002}$ \%, as well as with respect to \citet{biv21}'s CS abundance of $0.028\pm0.003$ \%. Based on its abundance upper limit of 0.07 \% \citep{biv21}, OCS is worth considering as another possible CS parent, but this species can be ruled out due to its photodissociation rate of $9.6\times10^{-5}$~s$^{-1}$ \citep{hue15}), which is too small to be compatible with the observed CS distribution. Other, larger carbon and sulfur-containing molecules should be considered as plausible candidates for the CS parent, such as CH$_3$SH, C$_2$H$_6$S, CH$_4$S$_2$ and C$_4$H$_6$S, which were detected using the ROSINA instrument at comet 67P \citep{cal16,han22}. However, since the photodissociation rates for these molecules are unknown, it remains to be seen if they could be consistent with the ALMA data. The difficultly of forming CS from any of these species in the inner coma should also be emphasized, due to the need to break multiple covalent bonds before CS can be liberated into the gas phase.

Further evidence against CS$_2$ as the parent of cometary CS was provided by the IRAM 30 m observations of comet 67P by \citet{biv23}, who obtained a CS abundance of $0.05 \pm 0.01$ \%. This is significantly larger than the CS$_2$ abundance measured by Rosetta during the previous apparition of 67P \citep[0.02 \%;][]{lau20}, so assuming the coma chemistry was the same on the two apparitions, an additional source of CS is required in that comet.

The identity of the main CS parent in comets therefore remains elusive. Considering our present knowledge, the most plausible sources include dust grains rich in carbon and sulfur, or possibly large C and S-bearing molecules (such as H$_n$C$_m$S$_k$). Thermal degradation of large molecules or dust could also explain the variation in CS/HCN ratio as a function of heliocentric distance observed by \citet{biv06,biv11}, assuming the CS production rate depends on the dust temperature. Future studies of the CS parent may benefit from the use of more physically realistic vectorial (or Monte Carlo) coma models, to account for the excess CS kinetic energy introduced during CS$_2$ photolysis, that might broaden its spatial distribution, leading to larger $L_p$ values. The possibility that the CS$_2$ photolysis lifetime is larger than the currently-accepted value should also be considered.

The CS production rate in the 46P coma ($0.022^{+0.004}_{-0.002}$ \% with respect to water), is at the lower end of the range of values previously found in comets (0.02–-0.20 \%; \citealt{boc17}). 

\subsection{HNC}

Our HNC parent scale length of $1412^{+14,876}_{-367}$~km in 46P is comparable with the values of $700^{+1,100}_{-400}$~km in S1/ISON \citep{cor14} and $3300^{+19,700}_{-2800}$~km in ER61 \citep{rot21b}. Unfortunately, the large error bars hinder the identification of any trends in $L_p({\rm HNC})$ with heliocentric distance or any other cometary parameters, so more observations at higher S/N will be needed to further elucidate the behavior of the HNC parent.

A strong variation in the HNC/HCN abundance ratio as a function of $r_H$ was observed in a sample of 14 moderately active comets by \citet{lis08}. This trend was interpreted as arising from variations in the HNC production rate as a function of coma dust grain temperature. Given the difficulty in producing HNC from known gas-phase chemical processes in the coma \citep{rod01}, it was therefore concluded that HNC is released from the thermal breakdown of macromolecules or polymeric material originating from inside the nucleus (see also \citealt{cor17a}). Our 46P HNC/HCN ratio of $5^{+19}_{-2}$ \% is in line with the observed trend with $r_H$, so similar breakdown of nitrogen-rich particles could be responsible for the observed HNC in comet 46P. The precise nature of those particles, however, remains to be elucidated.

\section{Conclusions}

ALMA observations of comet 46P provided an unusually close-up view of the coma, allowing quantitative measurements of molecular production scale lengths for the first time in a Jupiter-family comet. The interferometric data were analysed using a 3D radiative transfer model, consisting of a radially-expanding ambient coma and a broad, (near-)sunward jet of half-opening angle $\theta_{jet}=70^{\circ}$.   HCN is identified as a parent species, with a production scale length $L_p<3$~km (at 99\% confidence) and no evidence for any production of this molecule in the coma. CH$_3$OH is found to have comparable contributions from a nucleus source and a near-nucleus coma source (with $L_p=36\pm7$ km in the jet; $20\pm4$~km in the ambient coma). The CH$_3$OH coma source is explained as most likely originating from the sublimation of dirty ice grains or larger ice chunks in the coma, consistent with the hyperactive nature of this comet. The CH$_3$CN data are also consistent with a parent model, with an upper limit of $L_p<18$~km, demonstrating for the first time, that this species is most likely a parent molecule and originates from ices stored within the nucleus. We therefore conclude that measurements of HCN, CH$_3$OH and CH$_3$CN abundances in fully-activated comets provide a valid proxy for the abundances of these molecules within cometary nuclei.

The H$_2$CO, CS and HNC observations, on the other hand, cannot be reproduced using a parent model. Our models show that these species originate in the coma, either as photochemical daughters, or from the breakdown of macromolecules or dust grains, with $L_p$ values in the range 550--16,000~km. The detection of a distributed H$_2$CO source with $L_p=876^{+250}_{-175}$~km, combined with a lack of any detectable nucleus (parent) source for this molecule, is consistent with the non-detection of H$_2$CO from the 46P nucleus at infrared wavelengths.

Additional, spatial mapping of the distributed/daughter molecules in different comets (at various heliocentric distances), combined with laboratory studies of the dissociation rates of their putative parent species will be required to conclusively identify the parent materials. Given the consistency of the measured production scale lengths for H$_2$CO and HNC with previous ALMA observations of Oort cloud comets, our observations suggest that similar parent materials are present in JFCs and OCCs. Such parent materials must therefore be resistant to the thermal processing experienced by JFCs during their repeated perihelion passages.

The observed spectral and spatial data are consistent with an asymmetric coma, and we find enhanced outflow velocities and production rates for all species on the sunward (day side) of the comet. Conversely, the coma kinetic temperature is found to be significantly lower on the day side than the night side, which may be a result of enhanced adiabatic cooling rates on the day side.

The abundances of HCN and CH$_3$OH in 46P are consistent with average values found in OCCs and JFCs. On the other hand, CH$_3$CN, H$_2$CO and CS are at the lower end of the range of previously observed abundances, consistent with a depletion of these species (or their parents) in the 46P nucleus. We find that the abundances measured for daughter species (H$_2$CO, HNC and CS) are strongly dependent on the assumed parent scale length ($L_p$). We therefore emphasize the importance of using the correct $L_p$ value when deriving the abundances of cometary daughter/product species, particularly for comets at relatively small geocentric distances, thus demonstrating a crucial need for further spatially resolved studies of cometary molecular emission.

\begin{acknowledgments}
This work was supported by the National Science Foundation under Grant No. AST-2009253 (MAC) and AST-2009398 (BPB). The work of MAC, SNM, NXR and SBC was also supported by NASA's Planetary Science Division Internal Scientist Funding Program through the Fundamental Laboratory Research work package (FLaRe). This work makes use of ALMA dataset ADS/JAO.ALMA\#2018.1.01114.S. Part of this research was carried out at the Jet Propulsion Laboratory, California Institute of Technology, under NASA contract 80NM0018D0004. ALMA is a partnership of ESO (representing its member states), NSF (USA) and NINS (Japan), together with NRC (Canada) and NSC and ASIAA (Taiwan), in cooperation with the Republic of Chile. The Joint ALMA Observatory is operated by ESO, AUI/NRAO and NAOJ. The National Radio Astronomy Observatory is a facility of the National Science Foundation operated under cooperative agreement by Associated Universities, Inc.

\end{acknowledgments}

\appendix

\bibliography{refs}{}
\bibliographystyle{aasjournal}

\section{SUBLIME model geomtery}
\label{sec:geom}

\begin{figure}
\centering
\includegraphics[width=0.6\textwidth]{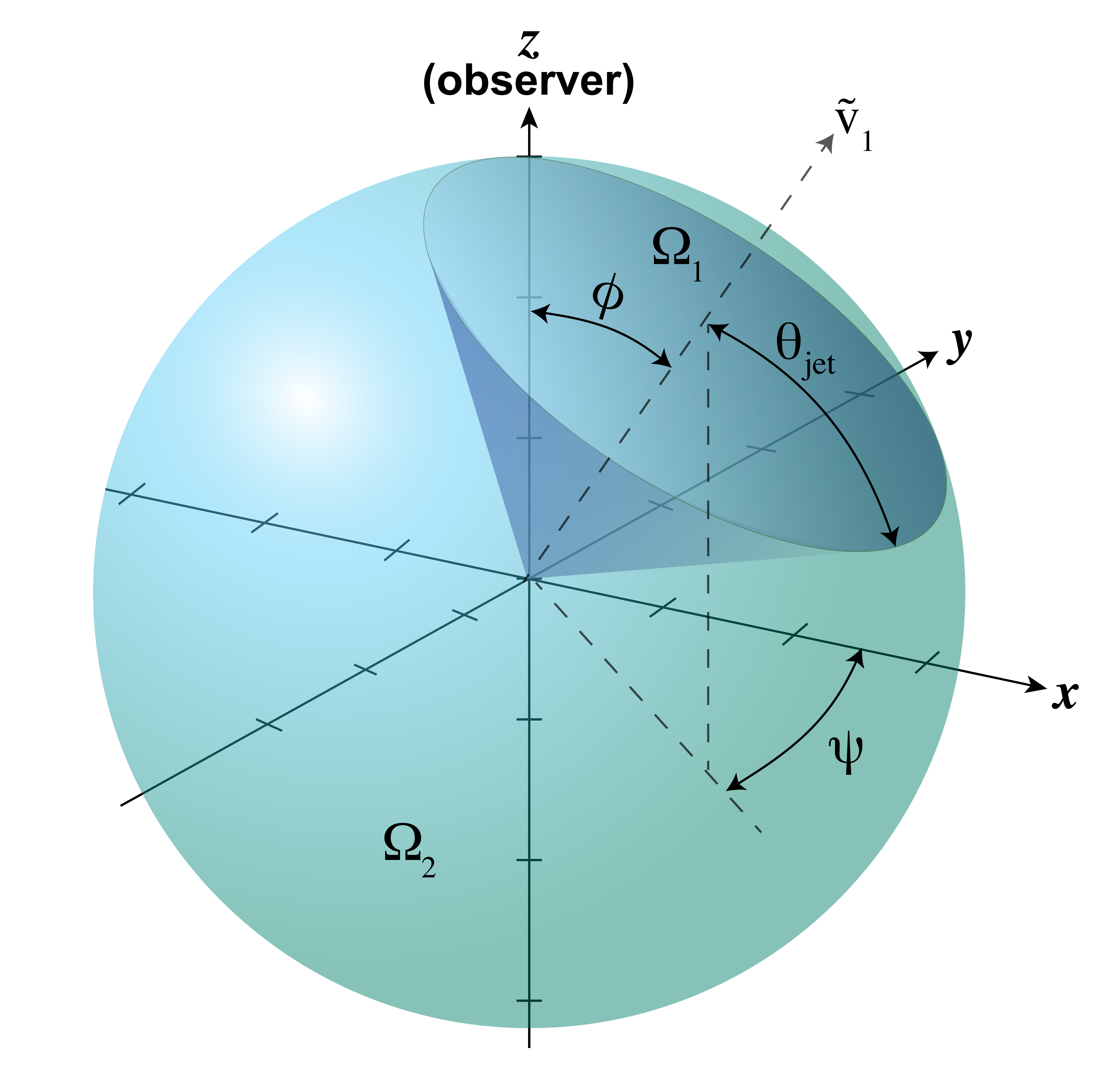}
\caption{SUBLIME model geometry used in the present study, showing the conical jet ($\Omega_1$) and ambient coma ($\Omega_2$), in cartesian coordiantes. $\theta_{jet}$ is the half-opening angle of the conical jet, $\phi$ is the (phase) angle of the jet axis with respect to the observer ($z$), and $\psi$ is the jet position angle in the plane of the sky ($x$ is north, $y$ is east).   \label{fig:geom}}
\end{figure}

Figure \ref{fig:geom} shows a schematic diagram of the two-component coma radiative transfer model geometry used in the present study. 

\section{Interferometric Visibility Spectral Modeling}
\label{sec:visfits}

\begin{figure}
\centering
\includegraphics[width=1.0\columnwidth]{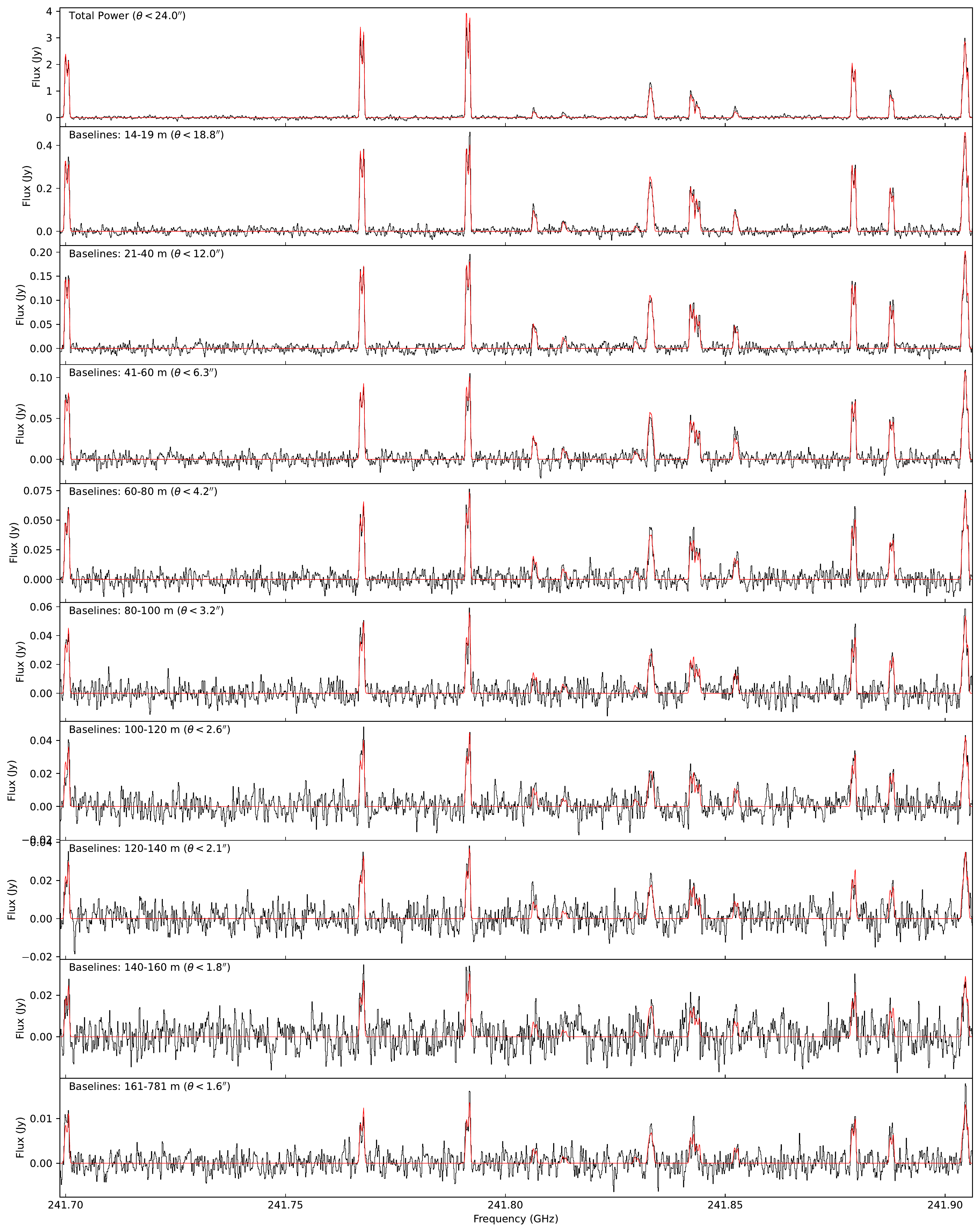} 
\caption{CH$_3$OH $J_K=5_K-4_K$ binned visibility spectra (real component), with best-fitting model overlaid in red.  Each sub-panel shows the average spectrum for a range of $uv$ distances (antenna baselines), given upper-left, which correspond to a range of angular scales ($\theta$) on the sky. \label{fig:ch3ohstack}}
\end{figure}

\begin{figure}
\centering
\includegraphics[width=0.5\columnwidth]{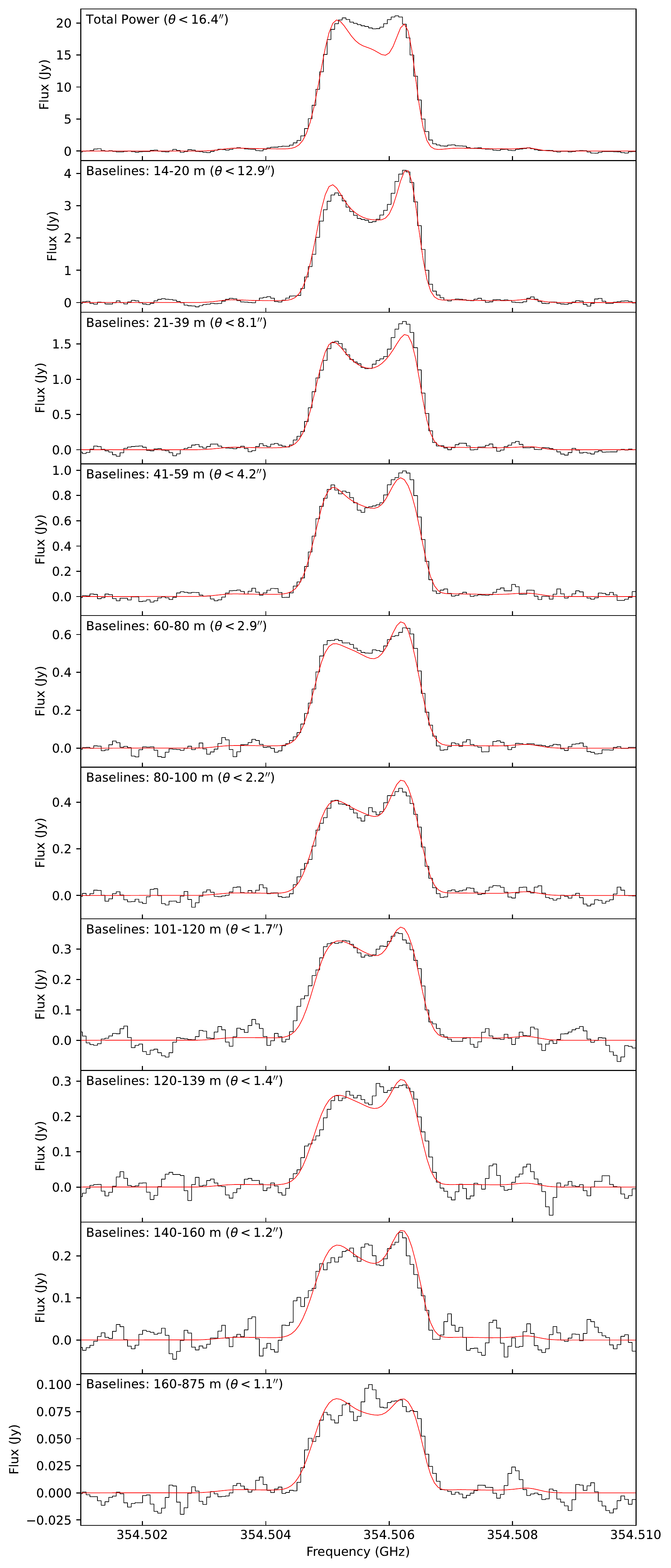} 
\caption{HCN $J=4-3$ binned visibility spectra (real component), with best-fitting model overlaid in red. Each sub-panel shows the average spectrum for a range of $uv$ distances (antenna baselines), given upper-left, which correspond to a range of angular scales ($\theta$) on the sky.  \label{fig:hcnstack}}
\end{figure}

\begin{figure}
\centering
\includegraphics[width=0.5\columnwidth]{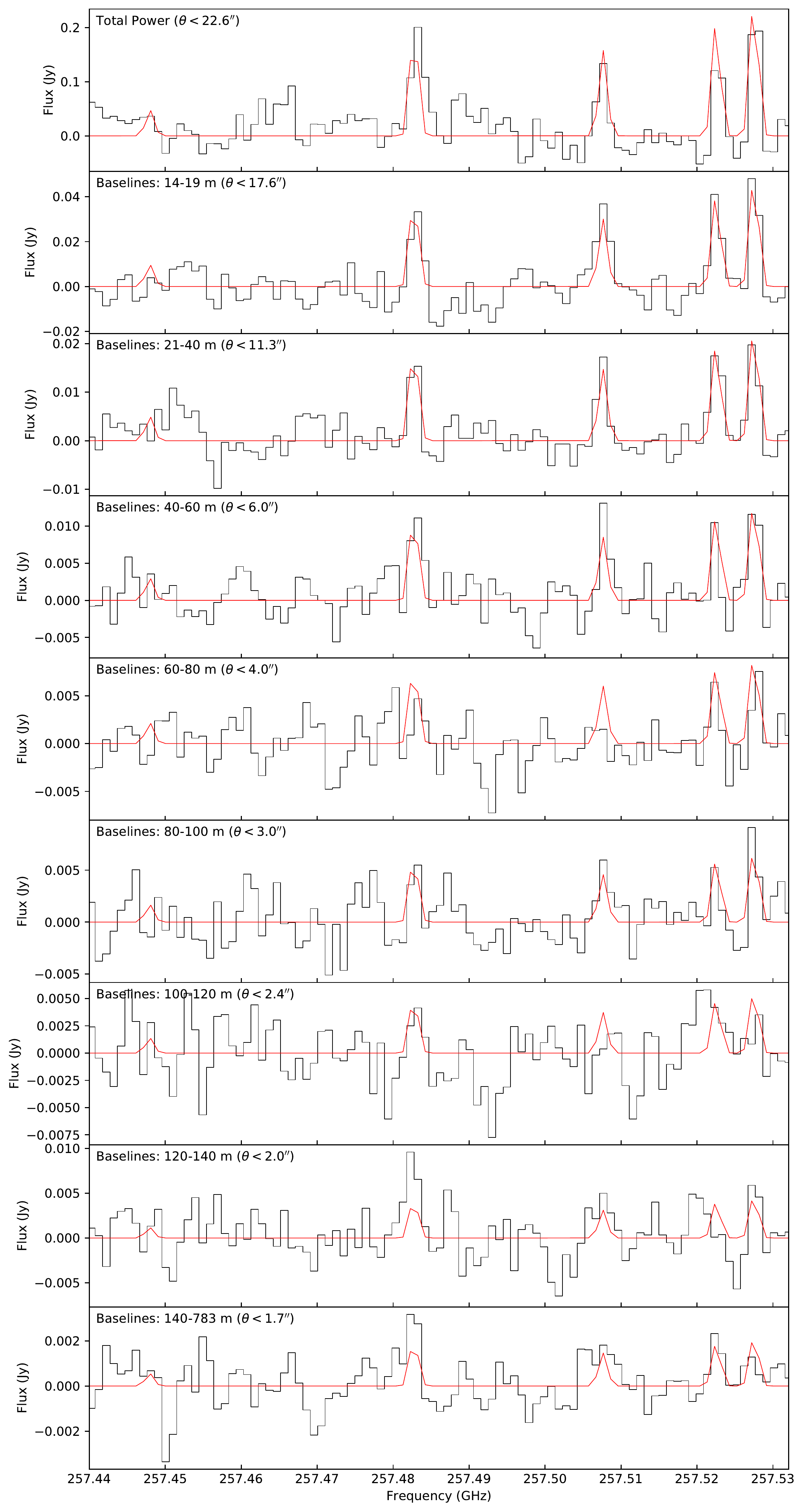} 
\caption{CH$_3$CN $J_K=14_K-13_K$ binned visibility spectra (real component), with best-fitting model overlaid in red. Each sub-panel shows the average spectrum for a range of $uv$ distances (antenna baselines), given upper-left, which correspond to a range of angular scales ($\theta$) on the sky.  \label{fig:ch3cnstack}}
\end{figure}

\begin{figure}
\centering
\includegraphics[width=0.5\columnwidth]{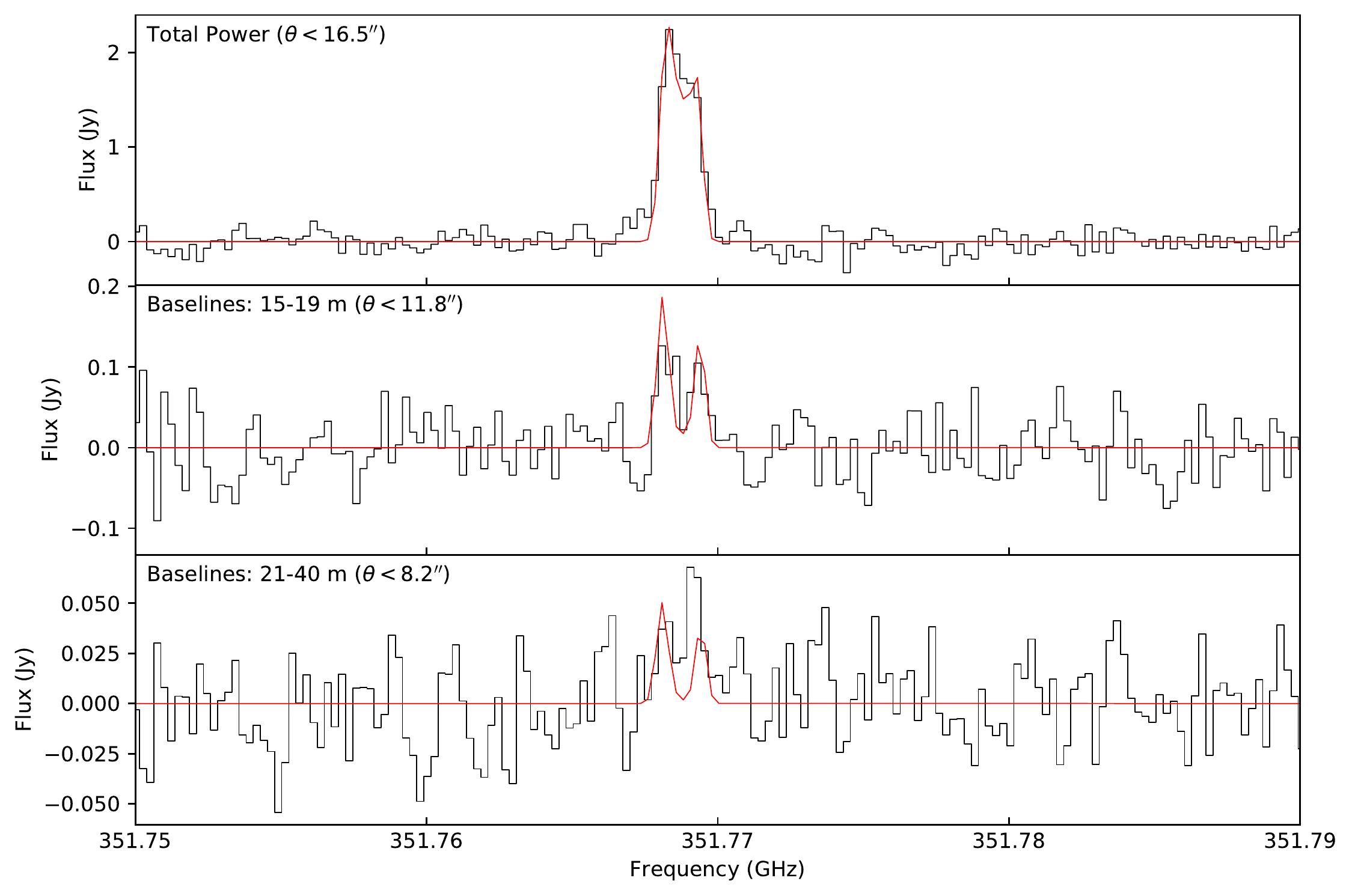} 
\caption{H$_2$CO $J_{K_a,K_c}=5_{1,5}$--$4_{1,4}$ binned visibility spectra (real component), with best-fitting model overlaid in red. Each sub-panel shows the average spectrum for a range of $uv$ distances (antenna baselines), given upper-left, which correspond to a range of angular scales ($\theta$) on the sky.  \label{fig:h2costack}}
\end{figure}

\begin{figure}
\centering
\includegraphics[width=0.5\columnwidth]{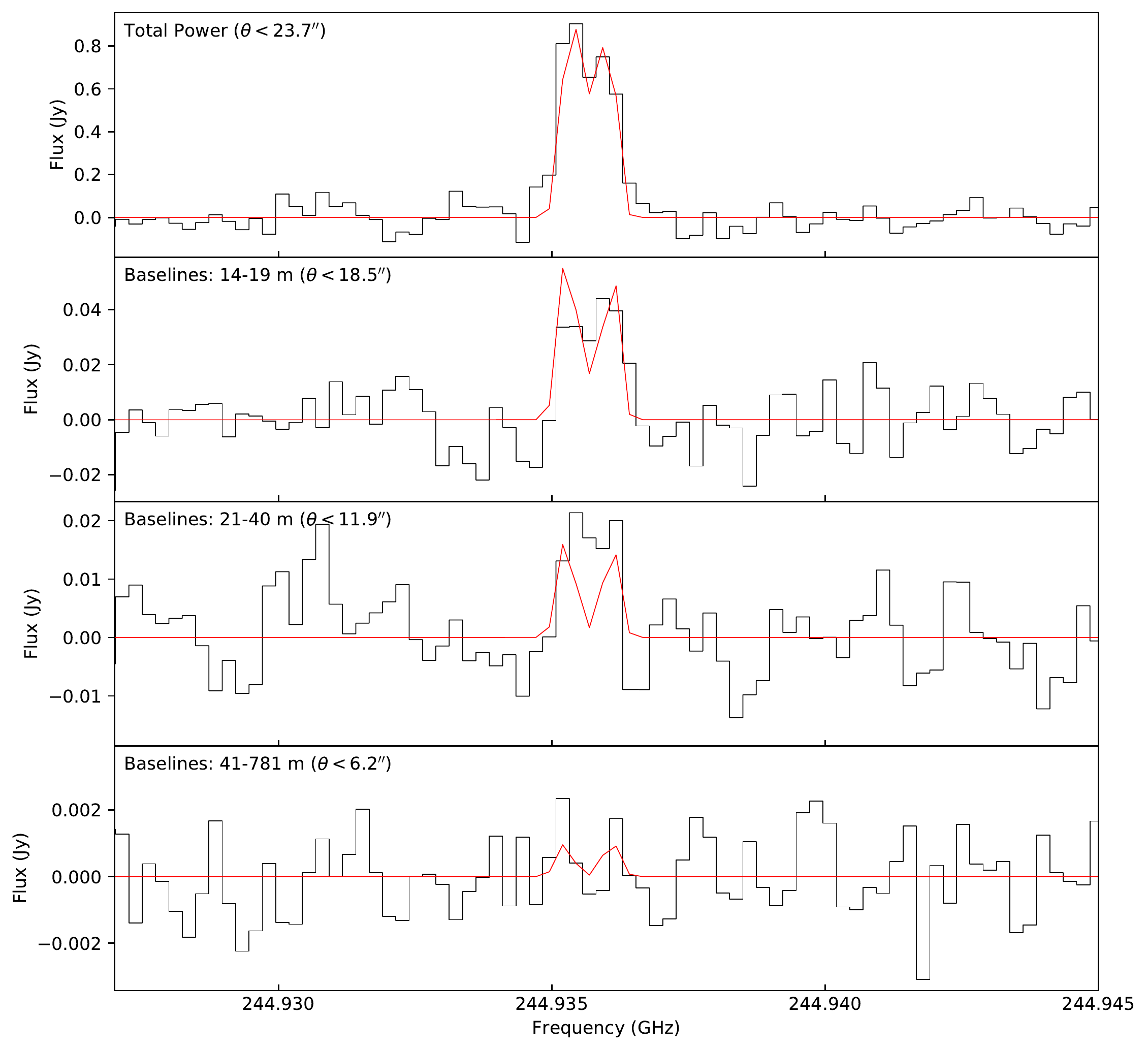} 
\caption{CS $J=5$--$4$ binned visibility spectra (real component), with best-fitting model overlaid in red. Each sub-panel shows the average spectrum for a range of $uv$ distances (antenna baselines), given upper-left, which correspond to a range of angular scales ($\theta$) on the sky.  \label{fig:csstack}}
\end{figure}

\begin{figure}
\centering
\includegraphics[width=0.5\columnwidth]{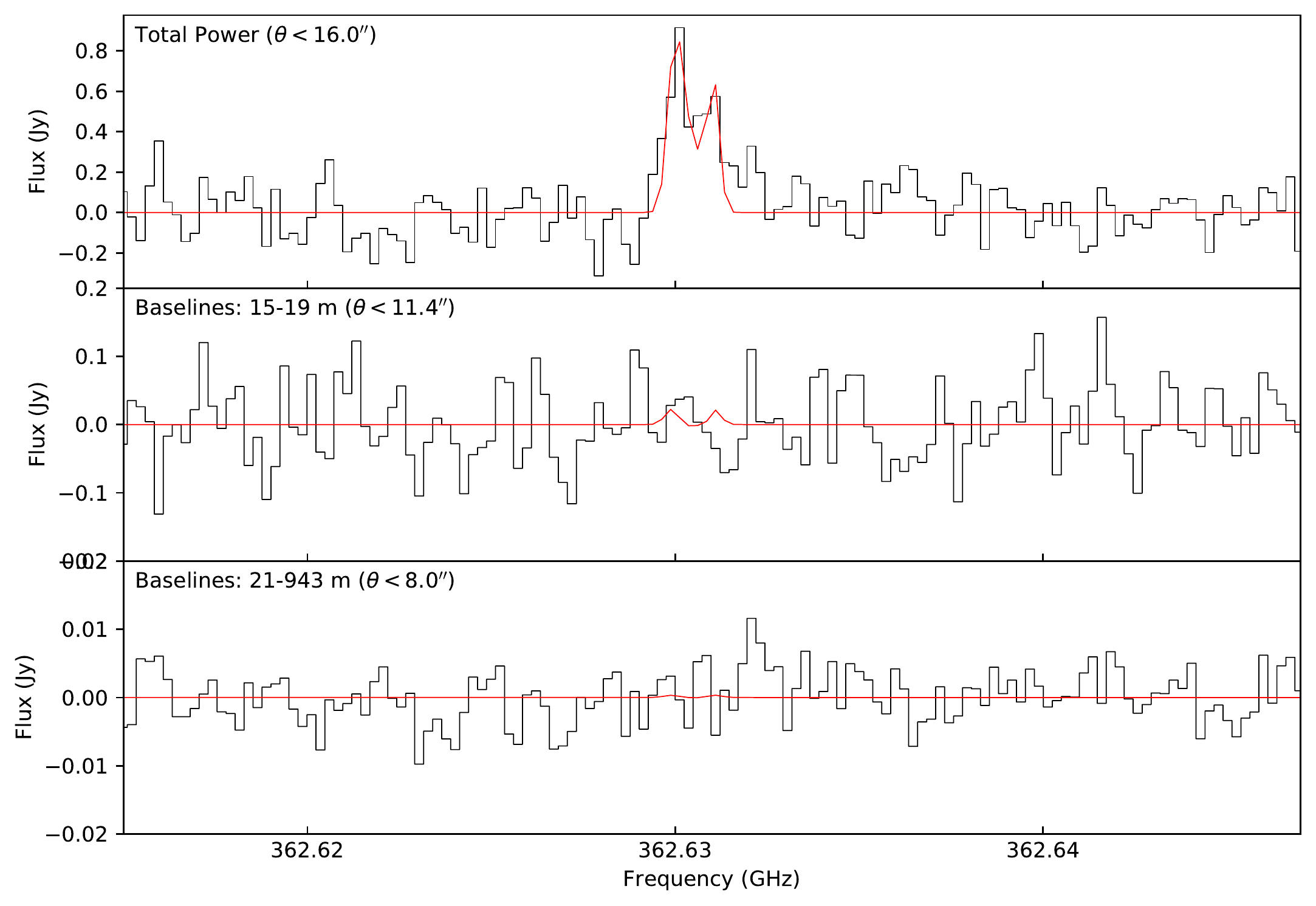} 
\caption{HNC $J=4$--$3$ binned visibility spectra (real component), with best-fitting model overlaid in red. Each sub-panel shows the average spectrum for a range of $uv$ distances (antenna baselines), given upper-left, which correspond to a range of angular scales ($\theta$) on the sky.  \label{fig:hncstack}}
\end{figure}

\begin{figure}
\centering
\includegraphics[width=0.6\textwidth]{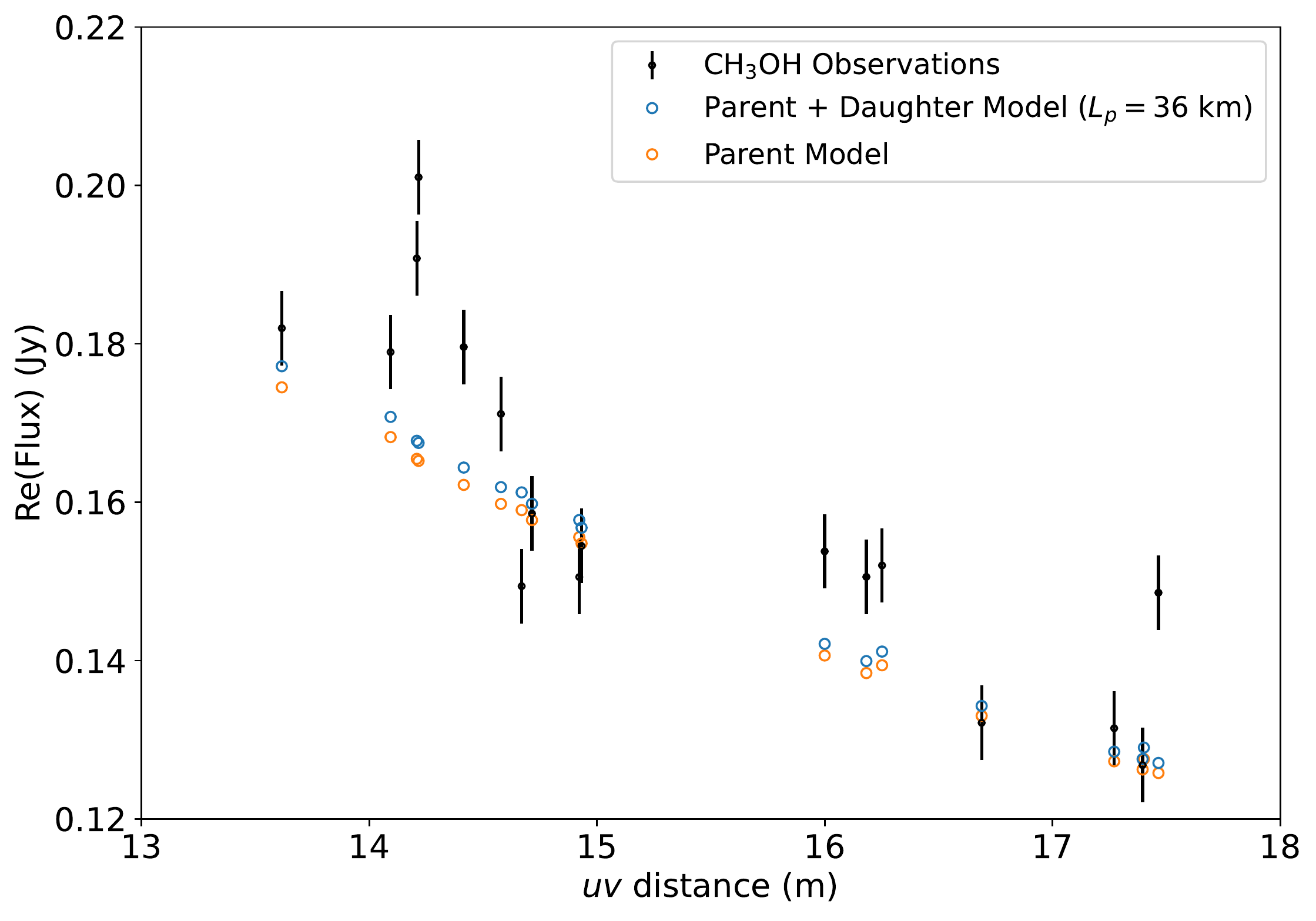}
\caption{Observed, spectrally averaged real interferometric visibilities for CH$_3$OH, zoomed in on the region of shortest baselines to show the improved fit of the parent + daughter model (blue circles) compared with the pure parent model (orange circles). \label{fig:ch3ohviszoom}}
\end{figure}

Figures \ref{fig:ch3ohstack} to \ref{fig:hncstack} show the real part of the observed and modeled ALMA visibilities for comet 46P, as a function of spectral channel.  The top panel of each figure shows the autocorrelation (total power) spectrum, while each successive sub-panel shows the data averaged over an range of increasing antenna baseline lengths (corresponding to decreasing angular scales). Figure \ref{fig:ch3ohviszoom} shows a zoomed, un-binned version of Figure \ref{fig:parentVis} for CH$_3$OH, to highlight the difference between the two best-fitting models and the data.

\end{document}